\documentclass[a4paper,english,cleveref]{lipics-v2019}

\nolinenumbers

\title{The Time-Triggered Wireless Architecture} 

\author{Romain Jacob}%
{ETH Zurich, Switzerland}
{}
{https://orcid.org/0000-0002-2218-5750}
{} 
\author{Licong Zhang}%
{TU Munich, Germany}
{}
{}
{} 
\author{Marco Zimmerling}%
{TU Dresden, Germany}
{}
{https://orcid.org/0000-0003-1450-2506}
{German Research Foundation (DFG) within the Emmy Noether project NextIoT (grant ZI 1635/2-1)} 
\author{Jan Beutel}%
{ETH Zurich, Switzerland}
{}
{https://orcid.org/0000-0003-0879-2455}
{} 
\author{Samarjit Chakraborty}%
{University of North Carolina at Chapel Hill, United States}
{}
{https://orcid.org/0000-0002-0503-6235}
{} 
\author{Lothar Thiele}%
{ETH Zurich, Switzerland}
{}
{https://orcid.org/0000-0001-6139-868X}
{Swiss National Science Foundation program "NCCR Automation"} 

\authorrunning{R. Jacob, L. Zhang, M. Zimmerling, J. Beutel, S. Chakraborty, L. Thiele}

\Copyright{Romain Jacob, Licong Zhang, Marco Zimmerling, Jan Beutel, Samarjit Chakraborty, and Lothar Thiele}



\category{}

\relatedversion{Prior 4-page publication~\cite{jacob2018ttw}: \url{https://doi.org/10.23919/DATE.2018.8342127}}

\supplement{%
\href{https://ttw.ethz.ch}{Project webpage} --
\href{https://github.com/romain-jacob/TTW-Artifacts}{TTW Artifacts} --
\href{https://github.com/romain-jacob/TTW-Artifacts/blob/master/resources/CRediT.pdf}{Authors' Contributions (CRediT)}%
}

\hideLIPIcs  

\EventEditors{Marcus V\"{o}lp}
\EventNoEds{1}
\EventLongTitle{32nd Euromicro Conference on Real-Time Systems (ECRTS 2020)}
\EventShortTitle{ECRTS 2020}
\EventAcronym{ECRTS}
\EventYear{2020}
\EventDate{July 7--10, 2020}
\EventLocation{Virtual Conference}
\EventLogo{}
\SeriesVolume{165}
\ArticleNo{19}

\usepackage{microtype}
\usepackage{xspace}
\usepackage{stackengine}

\usepackage{xcolor}
\definecolor{orange}{RGB}{255,153,51}
\definecolor{lightorange}{RGB}{255,235,214}
\definecolor{darkorange}{RGB}{171,68,1}
\definecolor{lightgrey}{RGB}{242,242,242}
\definecolor{midgrey}{RGB}{191,191,191}

\newcommand{\startsquarepar}{%
    \par\begingroup \parfillskip 0pt \relax}
\newcommand{\stopsquarepar}{%
    \par\endgroup}
\newcommand{\squarepar}[1]{%
\startsquarepar
#1
\stopsquarepar
}

\theoremstyle{definition}

\usepackage{booktabs}
\usepackage{mathtools}
\makeatletter
\newcommand{\ssymbol}[1]{$^{\@fnsymbol{#1}}$}
\makeatother

\usepackage{graphicx}
\graphicspath{{./}
{./Figures/}}
\DeclareGraphicsExtensions{.pdf,.jpeg,.png,.jpg}

\usepackage{afterpage}

\usepackage{relsize}

\usepackage{appendix}

\usepackage[ruled]{algorithm}
\usepackage{algpseudocode}

\crefname{section}{§}{§§}

\newcommand{\fakepar}[1]{\vspace{1mm}\noindent\textbf{#1.}}

\newcommand{\eg}{\emph{e.g.},\xspace}

\newcommand{\ie}{\emph{i.e.},\xspace}

\newcommand{\capt}[1]{\mdseries{\emph{#1}}}

\newcommand{\baloo}{Baloo\xspace}
\newcommand{\triscale}{TriScale\xspace}
\newcommand{\TriScale}{\triscale}
\newcommand{\DRP}{DRP\xspace}

\newcommand{\bolt}{Bolt\xspace}
\newcommand{\DPP}{DPP\xspace}

\newcommand{\TTW}{TTW\xspace}
\newcommand{\TTWlong}{Time-Triggered Wireless\xspace}
\newcommand{\TTnet}{TTnet\xspace}

\newcommand{\cps}{CPS\xspace}
\newcommand{\CPS}{CPS\xspace}

\newcommand{\customBox}[1]{%
\makebox[30pt][l]{#1}}

\newcommand{\nslotsmax}{\ensuremath{B_{\mathit{max}}}\xspace}
\newcommand{\nodeset}{\ensuremath{\mathcal{N}}\xspace}

\newcommand{\appset}{\ensuremath{\mathcal{A}}\xspace}
\newcommand{\persappset}{\ensuremath{\mathcal{A}_{P}}\xspace}
\newcommand{\taskset}{\ensuremath{\mathcal{T}}\xspace}
\newcommand{\messageset}{\ensuremath{\mathcal{M}}\xspace}
\newcommand{\modeset}{\ensuremath{\mathcal{O}}\xspace}
\newcommand{\predG}{\ensuremath{\mathbb{P}}\xspace}

\newcommand{\sched}[1]{\ensuremath{Sched(#1)}\xspace}

\newcommand{\objective}[1]{{\bf(O{#1})\xspace}}

\newcommand{\objB}{\objective{2}\xspace}
\newcommand{\constraint}[1]{{\bf(C{#1})\xspace}}

\newcommand{\appl}[1]{\ensuremath{\mathit{a_{#1}}}\xspace}

\newcommand{\app}{\appl{}}
\newcommand{\appA}{\ensuremath{\mathit{A}}\xspace}
\newcommand{\appB}{\ensuremath{\mathit{B}}\xspace}
\newcommand{\appX}{\ensuremath{\mathit{X}}\xspace}

\newcommand{\knownApp}[1]{\ensuremath{\mathit{K_{#1}}}\xspace}
\newcommand{\knownAppi}{\knownApp{i}}
\newcommand{\freeApp}[1]{\ensuremath{\mathit{F_{#1}}}\xspace}
\newcommand{\freeAppi}{\freeApp{i}}
\newcommand{\legApp}[1]{\ensuremath{\mathit{L_{#1}}}\xspace}
\newcommand{\legAppi}{\legApp{i}}
\newcommand{\legAppj}{\legApp{j}}
\newcommand{\virtlegApp}[1]{\ensuremath{\mathit{VL_{#1}}}\xspace}
\newcommand{\virtlegAppi}{\virtlegApp{i}}

\newcommand{\minminvirtlegApp}[2]{\ensuremath{\widehat{\virtlegApp{#1}^{#2}}}\xspace}
\newcommand{\minvirtlegApp}[2]{\ensuremath{\widetilde{\virtlegApp{#1}^{#2}}}\xspace}

\newcommand{\cf}[1]{\ensuremath{\mathit{CF({#1})}}\xspace}

\newcommand{\mode}[1]{\ensuremath{\mathit{M_{#1}}}\xspace}
\newcommand{\modeany}{\mode{}}
\newcommand{\modei}{\mode{i}}
\newcommand{\modej}{\mode{j}}
\newcommand{\modeGraph}{\ensuremath{\mathbb{M}}\xspace}
\newcommand{\modeGraphA}{\ensuremath{\mathbb{G}_A}\xspace}
\newcommand{\modeHyperperiod}{\ensuremath{\mathit{LCM}}\xspace}

\newcommand{\Tround}{\ensuremath{T_{r}}\xspace}
\newcommand{\Troundj}{\ensuremath{T_{r_j}}\xspace}
\newcommand{\Troundk}{\ensuremath{T_{r_k}}\xspace}

\newcommand{\Toffset}{\ensuremath{T_{o}}\xspace}
\newcommand{\Troundon}{\ensuremath{T_{r}^{\mathit{on}}}\xspace}
\newcommand{\Tworound}{\ensuremath{T_{\mathit{wo/r}}}\xspace}
\newcommand{\Tworoundon}{\ensuremath{T_{\mathit{wo/r}}^{\mathit{on}}}\xspace}

\newcommand{\Tslot}{\ensuremath{T_{\mathit{slot}}}\xspace}
\newcommand{\Twakeup}{\ensuremath{T_{\mathit{wake-up}}}\xspace}
\newcommand{\Tstart}{\ensuremath{T_{\mathit{start}}}\xspace}
\newcommand{\Tglossy}{\ensuremath{T_{\mathit{flood}}}\xspace}

\newcommand{\Thop}{\ensuremath{T_{\mathit{hop}}}\xspace}
\newcommand{\Td}{\ensuremath{T_{d}}\xspace}
\newcommand{\Tcal}{\ensuremath{T_{\mathit{cal}}}\xspace}
\newcommand{\Lcal}{\ensuremath{L_{\mathit{cal}}}\xspace}
\newcommand{\Theader}{\ensuremath{T_{\mathit{header}}}\xspace}
\newcommand{\Lheader}{\ensuremath{L_{\mathit{header}}}\xspace}
\newcommand{\Tpayload}{\ensuremath{T_{\mathit{payload}}}\xspace}
\newcommand{\Rbit}{\ensuremath{R_{\mathit{bit}}}\xspace}
\newcommand{\Tgap}{\ensuremath{T_{\mathit{gap}}}\xspace}
\newcommand{\Tpreprocess}{\ensuremath{T_{\mathit{preprocess}}}\xspace}
\newcommand{\Ton}{\ensuremath{T^{\mathit{on}}}\xspace}
\newcommand{\Toff}{\ensuremath{T^{\mathit{off}}}\xspace}
\newcommand{\Lbeacon}{\ensuremath{L_{\mathit{beacon}}}\xspace}
\newcommand{\Lmax}{\ensuremath{L_{\mathit{max}}}\xspace}
\newcommand{\nslotsround}{\ensuremath{B_r}\xspace}
\newcommand{\nslots}{\ensuremath{B}\xspace}

\newcommand{\map}{\ensuremath{\mathit{map}}\xspace}
\newcommand{\prio}{\ensuremath{\mathit{prio}}\xspace}

\newcommand{\af}{\ensuremath{\mathit{af}}\xspace}
\newcommand{\df}{\ensuremath{\mathit{df}}\xspace}
\renewcommand{\sf}{\ensuremath{\mathit{sf}}\xspace}
\newcommand{\id}{\ensuremath{\mathit{id}}\xspace}
\newcommand{\ids}{\ensuremath{\mathit{id}}s\xspace}

\newcommand{\s}{\ensuremath{\,\text{s}}\xspace}
\newcommand{\ms}{\ensuremath{\,\text{ms}}\xspace}
\newcommand{\us}{\ensuremath{\,\mu\text{s}}\xspace}

\newcommand{\kHz}{\ensuremath{\,\text{kHz}}\xspace}
\newcommand{\MHz}{\ensuremath{\,\text{MHz}}\xspace}

\newcommand{\kbps}{\ensuremath{\,\text{kbps}}\xspace}

\newcommand{\bytes}{\ensuremath{~\text{bytes}}\xspace}

\usepackage{ifthen}
\newboolean{authnotes}
\newboolean{markfinalchanges}

\setboolean{authnotes}{true}

\ifthenelse{\boolean{authnotes}}
{
\newcommand{\rj}[1]{\footnote{{\bf RJ: #1}}}
\newcommand{\cb}[1]{\footnote{{\bf CB: #1}}}
\newcommand{\lv}[1]{\footnote{{\bf LV: #1}}}
\newcommand{\lt}[1]{\footnote{{\bf LT: #1}}}
\newcommand{\mz}[1]{\footnote{{\bf MZ: #1}}}
}
{
\newcommand{\rj}[1]{}
\newcommand{\cb}[1]{}
\newcommand{\lv}[1]{}
\newcommand{\lt}[1]{}
\newcommand{\mz}[1]{}
}

\setboolean{markfinalchanges}{false}

\ifthenelse{\boolean{markfinalchanges}}
{
\newcommand{\final}[1]{{\color{red} #1}}
}{
\newcommand{\final}[1]{#1}
}

\newcommand{\onlineroot}{https://nbviewer.jupyter.org/github/romain-jacob/TTW-Artifacts/blob/master}
\newcommand{\ttwfig}[1]{\onlineroot/ttw_plots.ipynb\#{#1}}

\newcommand{\TablePath}{./Tables}

\begin{document}

\maketitle


\begin{abstract}
Wirelessly interconnected sensors, actuators, and controllers promise greater flexibility, lower installation and maintenance costs, and higher robustness in harsh conditions than wired solutions.
However, to facilitate the adoption of wireless communication in cyber-physical systems (\CPS), the functional and non-functional properties must be similar to those known from wired architectures.
We thus present \TTWlong (\TTW), a wireless architecture for multi-mode \CPS that offers reliable communication with guarantees on end-to-end delays among distributed applications executing on low-cost, low-power embedded devices.
We achieve this by exploiting the high reliability and deterministic behavior of a synchronous transmission based communication stack we design, and by coupling the timings of distributed task executions and message exchanges across the wireless network by solving a novel co-scheduling problem.
While some of the concepts in \TTW have existed for some time and \TTW has already been successfully applied for feedback control and coordination of multiple mechanical systems with closed-loop stability guarantees, this paper presents the key algorithmic, scheduling, and networking mechanisms behind \TTW, along with their experimental evaluation, which have not been known so far.
%
%
\TTW is open source and ready to use: \href {https://ttw.ethz.ch}{\path{ttw.ethz.ch}}
\end{abstract}



\section{Introduction}
\label{sec:ttw_intro}

\squarepar{%
For decades, real-time distributed control systems have \final{mostly} relied on fieldbuses like CAN, FlexRay, and PROFIBUS.
Following the design principles of the Time-Triggered Architecture~(TTA)~\cite{kopetz2003TTA}, these systems provide predictability of functional and non-functional properties.
Yet, wired systems are increasingly reaching their limits as future cyber-physical systems (\CPS) demand higher flexibility and cost efficiency.
Low-power wireless communication promises to meet these demands by allowing for an unprecedented degree of mobility and deployment flexibility, avoiding cable breaks or faulty connections, and providing greater robustness to heat, humidity, abrasive substances, and undamped vibrations.
However, to be viable for mission-critical \cps, a wireless system must feature timing predictability similar to traditional wired systems.
Moreover, short end-to-end delays (tens of milliseconds), high reliability, and multiple years of battery lifetime are required for many applications~\cite{akerberg2011Future}.}

\squarepar{%
The past years have seen significant progress in this direction.
In particular, the concept of \emph{synchronous transmissions}~\cite{ferrari2011Glossy} has enabled highly reliable and efficient low-power wireless communication protocols that are robust to the unpredictable dynamics of wireless systems.
As further detailed in~\cite{zimmerling2020Survey}, this is because of two main reasons: (\emph{i}) synchronous transmissions enable the design of protocols whose logic is \emph{independent of the time-varying network state}, which leads to a highly deterministic protocol execution regardless of changes in the network; (\emph{ii}) synchronous-transmission-based communication protocols inherently exploit \emph{different forms of diversity}, \final{including} sender and receiver diversity as well as temporal and spatial diversity, \final{leading to a reliability as high as 99.9999\,\% in certain scenarios}~\cite{ferrari2011Glossy}.
In fact, since its inception in 2016, all top three teams at the annual EWSN Dependability Competition have built on synchronous transmissions, demonstrating dependable multi-hop communication even in extreme wireless interference scenarios~\cite{schuss2017Competition}.
Building on this dependable base, protocols have been designed \final{that} provide sub-microsecond network-wide time synchronization \final{while abstracting} from the complexity of the underlying \final{dynamic} network topology, thus allowing to reason about a low-power wireless network as if it were a shared bus~\cite{ferrari2012LWB}.%
}

\squarepar{%
In light of this recent progress, we ask: \emph{Is it possible to design a reliable, adaptive, and efficient time-triggered architecture for low-power wireless multi-hop networks with formal guarantees on end-to-end delays among distributed application tasks?}
If so, it would take wireless systems a decisive step closer to wired systems by providing similar abstractions and guarantees as those researchers, engineers, and operators are used to, thus enabling powerful new \CPS applications in industry, healthcare, energy, and many other domains.}

\fakepar{Challenges}
%
There are four key technical challenges standing in the way of a time-triggered architecture for low-power wireless multi-hop networks.

\begin{enumerate}
\squarepar{%

  \item \emph{Real-time communication in the face of radio duty cycling.}
  In a fieldbus, a node can listen idly without incurring costs, allowing it to react immediately to a request. In a low-power wireless bus~\cite{ferrari2012LWB}, instead, a node must turn its radio off whenever possible to save energy, which renders the node unreachable until the next scheduled wake-up.
  To reduce energy costs due to idle listening, wireless protocols schedule communication rounds, where all nodes \final{wake up}, exchange messages, and go back to sleep~(\eg \cite{ferrari2012LWB,wirelessHART,jacob2019Baloo,watteyne2017Teaching}).
  Deciding when rounds take place and which nodes can send messages in each round to meet real-time deadlines at minimum energy costs is a complex scheduling problem; for example, the allocation of messages to rounds resembles combinatorial NP-hard bin packing~\cite{korte2006Combinatorial}.

  \item \emph{Coupling of communication and application tasks.} What ultimately matters in a \CPS are the real-time constraints among application tasks executing on distributed devices.
  Hence, task executions and message exchanges over the wireless network must be coupled to guarantee end-to-end deadlines.
  A common approach for wired fieldbuses is to statically co-schedule all tasks and messages~\cite{abdelzaher1999combined, ashjaei2017Designing, craciunas2016Combined} to minimize delays by solving a satisfiability modulo theories~(SMT)~\cite{craciunas2014SMTbased,huang2012Static,steiner2010evaluation} or a mixed integer linear programming (MILP)~\cite{azim2014Scheduling} problem. To apply this approach to a wireless bus, we must embed the above-mentioned bin-packing problem into the schedule synthesis.
  This, however, introduces non-linear constraints \final{that} cannot be easily handled in a SMT or MILP formulation \final{(see \cref{sec:single_mode})}.

  \item \emph{Dependencies across modes.} Static co-scheduling lacks runtime adaptability.
  This is often mitigated by enabling the system to switch at runtime between multiple operation modes, each having a pre-computed scheduling table~\cite{fohler1993changing}.
  The real-time guarantees that can be provided across mode changes depend on the mode-change protocol~\cite{chen2018SafeMC} and how the modes are scheduled.
  For example, if the periodicity of a task should be preserved across a mode change, this task must have the same schedule in both modes.
  A na\"{i}ve approach to handle such dependencies across modes is a single MILP formulation with a global objective function.
  However, the poor scalability of the scheduling problem (NP-hard~\cite{jeffay1991nonpreemptive}) becomes a bottleneck for any realistic \CPS scenario with many modes, while holding no guarantee that the computed schedules perform efficiently in terms of energy.

  \item \emph{Worst-case communication time.} The synthesis of scheduling tables requires an accurate model of the timing of all operations, including the worst-case execution time of tasks \emph{and} the worst-case communication time of messages. While staple methods exist for tasks~\cite{wilhelm2008worstcase}, a predictable protocol implementation is needed that yields accurate upper bounds on the time required for exchanging messages across a multi-hop network.%
}
\end{enumerate}

\squarepar{%
\fakepar{Contributions}
This paper presents Time-Triggered Wireless (\TTW), a wireless architecture for multi-mode \CPS.
By addressing all of the above challenges, \TTW provides: (\emph{i})~highly reliable and efficient communication across dynamic low-power wireless multi-hop networks; (\emph{ii})~formal guarantees on end-to-end delays among distributed application tasks; (\emph{iii})~runtime adaptability through mode changes that respect the tasks' periodicity constraints.
With this, \TTW does not only significantly advance the state of the art in real-time wireless systems, but also represents a solid foundation for the design of dependable wireless \cps.}

\begin{figure}[!tb]
	\centering
    \stackinset{l}{250.5pt}{b}{4.5pt}{%
      \hyperref[subsec:ttnet_design]{%
      \makebox(18,12){}}}{%
    \stackinset{l}{157pt}{t}{88.5pt}{%
      \hyperref[sec:ttnet]{%
      \makebox(12,12){}}}{%
    \stackinset{l}{61pt}{t}{88.5pt}{%
      \hyperref[sec:scheduler]{%
      \makebox(12,12){}}}{%
    \stackinset{l}{61pt}{t}{54pt}{%
      \hyperref[subsec:ttnet_model]{%
      \makebox(18,12){}}}{%
    \stackinset{l}{1pt}{t}{39pt}{%
      \hyperref[subsec:systModel]{%
      \makebox(18,12){}}}{%
	   \includegraphics[width=1.0\linewidth]{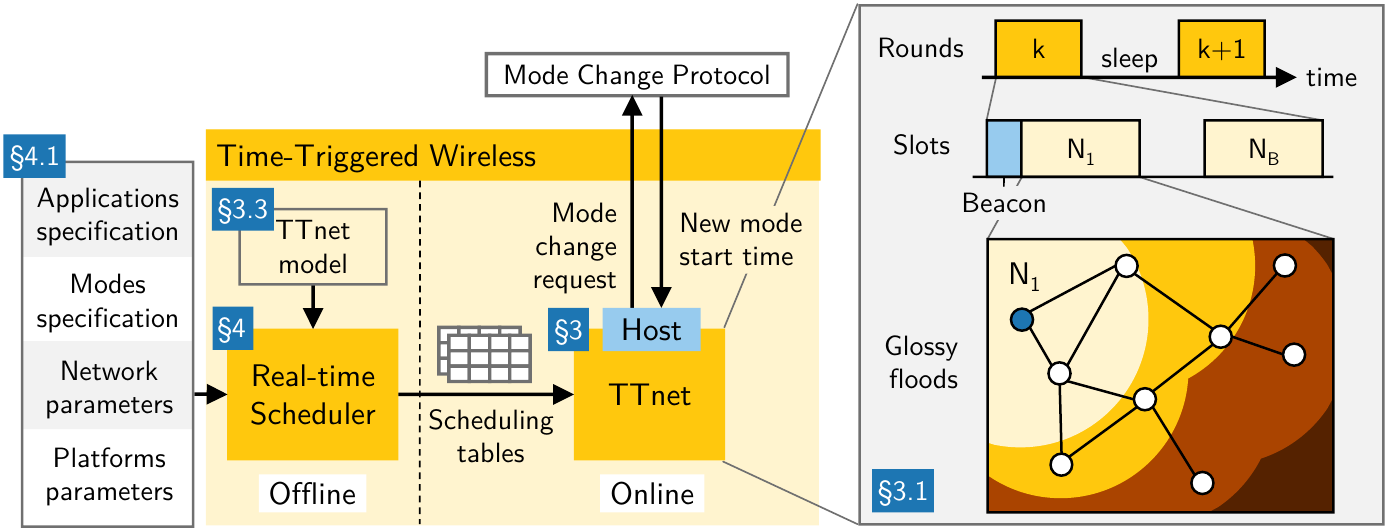}}}}}}
	\caption{
		High-level overview of the Time-Triggered Wireless (\TTW) architecture.
	 }
	 \vspace{-4mm}
	\label{fig:ttw_overview}
\end{figure}

As shown in \cref{fig:ttw_overview}, \TTW consists of two main components: a system-wide real-time scheduler that executes offline and a communication stack called \TTnet that runs online on distributed low-power wireless devices.
Based on the application specification (\eg tasks, messages, modes) and system parameters (\eg number of nodes, message sizes), the scheduler synthesizes optimized scheduling tables for the entire system.
These tables are loaded onto the devices, and at runtime each device follows the table that corresponds to the current mode.
\TTnet provides a generic interface to implement different mode-change protocols~\cite{chen2018SafeMC}.

\squarepar{%
Our design of the real-time scheduler uses concepts from network calculus~\cite{leboudec2001Network} and novel heuristics to addresses challenges \textsf{\textbf{1.}}--\textsf{\textbf{3.}}
Further, our design of \TTnet addresses challenge \textbf{\textsf{4.}} by exploiting synchronous transmissions, in particular the deterministic behavior of Glossy floods~\cite{ferrari2011Glossy}.
As indicated for node N$_{1}$ in the bottom right corner of \cref{fig:ttw_overview}, a Glossy flood sends a message to all nodes in a multi-hop network with a reliability that \final{can} exceed 99.9\,\% in \final{certain} networks and challenging conditions~\cite{ferrari2012LWB,schuss2017Competition}, yet the time this process takes can be confined to a time slot of known length.
\TTnet groups a series of such time slots into rounds to save energy.
This yields a highly timing-predictable protocol execution for which we devise timing and energy models as input for the \TTW real-time scheduler.

We implement \TTW on physical platforms and \final{open-source our implementation~\cite{repoTTW}.}
Using this implementation, we perform real-world experiments on 27 low-power wireless nodes of the FlockLab testbed~\cite{lim2013FlockLab}.
The results demonstrate that, for the settings we tested, our models are highly accurate and provide accurate upper bounds for the schedule synthesis; for example, our timing model overestimate the length of a round in \TTnet by at most 0.7\,\ms.
The results also show that our scheduling techniques can effectively reduce the energy cost of wireless communication in \TTW and make the scheduling problem tractable: the solving time for one mode on a standard laptop PC ranges from a few seconds to a few minutes depending on the complexity of the mode (\eg number of tasks and messages).

In addition, \TTW already proved its utility for practical wireless \CPS representative of emerging applications, such as remote control in chemical plants and cooperative robotics in smart manufacturing.
Specifically, \TTW was essential in enabling fast and reliable feedback control and coordination of multiple physical systems over low-power wireless multi-hop networks with closed-loop stability guarantees despite mode changes and mobile nodes~\cite{baumann2019TCPS, mager2019Feedback}.%
\footnote{\mbox{Public demo: \url{https://youtu.be/AtULmfGkVCE}. Mobility experiment: \url{https://youtu.be/19xPHjnobkY}.}}
\final{In these works, we integrated \TTW with a mode-change protocol and analyzed the worst-case end-to-end jitter; thanks to the predictability of \TTW, this jitter can be made negligible ($\leq 50\us$) for typical \CPS applications, which we empirically validated (see~\cite{baumann2019TCPS, mager2019Feedback} for details)}.%

\fakepar{Difference to prior paper}
This paper significantly extends a prior 4-page publication~\cite{jacob2018ttw} by (\emph{i}) incorporating mode changes in the system model and the formulation of the scheduling problem, (\emph{ii}) providing an implementation of \TTW on physical platforms, and (\emph{iii}) using this implementation to evaluate \TTW and validate our models through real-world experiments.%
}


\section{Related Work}
\label{sec:ttw_relWork}

\squarepar{%
\TTW is most closely related to prior work on reliable and predictable solutions for wireless sensor-\final{actuator} networks and real-ime distributed control systems based on fieldbuses.

In the wireless domain, numerous standards and protocols have been proposed for low-power multi-hop wireless networks, including WirelessHART, ISA100, and TSCH~\cite{watteyne2017Teaching} from industry and several proposals from academia (\eg~\cite{chipara2011InterferenceAware,saifullah2010RealTime,he2003SPEED}).
Closest to \TTW is Blink~\cite{zimmerling2017Blink}, which also builds on the concept of synchronous transmissions, in particular the Low-Power Wireless Bus~(LWB)~\cite{ferrari2012LWB}, to achieve adaptive real-time communication with guarantees on packet deadlines.
While certainly important and useful, the key difference to \TTW is that all these solutions consider only the network resources.
They do not take into account the scheduling of application tasks on the distributed devices, and can therefore not influence the end-to-end delays and jitter that ultimately \final{matter} from an application's perspective.
The only prior work that has looked at this end-to-end problem is DRP~\cite{jacob2016DRP}.
To provide end-to-end guarantees on packet deadlines, DRP decouples tasks and messages as much as possible, aiming for efficient support of sporadic or event-triggered applications.
Compared with \TTW, which tightly couples tasks and messages by co-scheduling them, \DRP incurs significantly higher worst-case delays, and is thus not suitable for demanding \cps applications~\cite{akerberg2011Future}.%
}

In the wired domain, a lot of work has been done on time-triggered architectures, such as the Time-Triggered Protocol (TTP)~\cite{kopetz1993TTP}, the static-segment of FlexRay~\cite{flexray2013ISO}, and Time-Triggered Ethernet~\cite{kopetz2005TimeTriggered}.
Like \TTW, many recent works in this area also use SMT or MILP based methods to synthesize and/or analyze static (co-)schedules for those architectures~\cite{ashjaei2017Designing,craciunas2016Combined,steiner2010evaluation,tamas2012Synthesis,zhang2014Task}.
The key difference to \TTW, however, is that these approaches assume that a message can be scheduled at any point in time.
\final{While this is a perfectly valid assumption for a wired system, it} is not compatible with the use of communication rounds in a wireless setting.
\cref{sec:ttw_evaluation_sched} shows that using  rounds greatly reduces the energy consumed for communication, but it makes the schedule synthesis more complex, as detailed in \cref{sec:scheduler}.


\section{Communication Stack}
\label{sec:ttnet}

\squarepar{%
We introduce \TTW, a time-triggered architecture that supports the design and implementation of dependable wireless \CPS.
This section presents the design, implementation, and \final{analytical models} of \TTW's communication stack \final{called} \TTnet.
\TTW's real-time scheduler, \final{described in} \cref{sec:scheduler}, uses these models to synthesize optimized scheduling tables.}

\subsection{\TTnet Design}
\label{subsec:ttnet_design}

\squarepar{%
To support a wide spectrum of \CPS applications possibly involving control and coordination of multiple physical systems over large distances, \TTW requires a low-power wireless stack that provides reliable many-to-all communication over multiple hops.
Further, the communication delays must be as short as possible and tightly bounded to support fast physical systems (\eg mechanical systems with dynamics of tens of milliseconds).
Finally, network-wide time synchronization is essential to reduce delays and achieve high performance and efficiency.

\TTnet addresses these requirements by taking inspiration from the Low-Power Wireless Bus~(LWB)~\cite{ferrari2012LWB} and \final{improving latency and efficiency by using a different scheduling strategy}.
The basic communication scheme is illustrated on the right side of \cref{fig:ttw_overview}, where distributed nodes are wirelessly interconnected and communicate using a sequence of Glossy floods~\cite{ferrari2011Glossy}.
As shown for when node N$_{1}$ sends its message, the flood spreads like a wave through the multi-hop network (note the different colors) so all nodes in the network can receive the message.
During the flood, nodes that receive the message at the same time also retransmit the message at the same time.
\final{This technique, known as synchronous transmissions,
is both
fast (it achieves the theoretical minimum latency for one-to-all flooding)
and
extremely reliable~\cite{zimmerling2020Survey}.}
Theoretical and empirical studies have shown that the few messages losses that do occur due to the vagaries of wireless communication are \final{largely} decorrelated, which greatly simplifies analytical modeling and control design~\cite{karschau2018Renormalization,zimmerling2013modeling}.
During a Glossy flood, the nodes also time-synchronize themselves with sub-microsecond accuracy~\cite{ferrari2011Glossy}. Finally, since the protocol logic is independent of the network topology\final{,} it is highly robust to changes inside the network (\eg due to mobile or failing devices) and external interference.

As shown in \cref{fig:ttw_overview}, \TTnet performs multiple Glossy floods in consecutive slots within a round; between two rounds all nodes have their wireless radio turned off to conserve energy.
The duration of the slots and rounds as well as their associated energy costs can be accurately modeled (\cref{subsec:ttnet_model}).
Which nodes get to send their messages in each slot is determined by \TTW's real-time scheduler.
The scheduler can abstract the entire multi-hop network with all its complexities and dynamics as a single shared resource with a known bandwidth, just like a wired fieldbus or a uniprocessor.
This is because \TTnet nodes appear to share a common clock and every messages is distributed to all nodes irrespective of the network topology.

\TTnet exploits the fact that the schedules are computed offline to minimize communication delays.
Instead of letting a dedicated host node compute the schedules between rounds and distribute them at the beginning of each round as in LWB, the host in \TTnet only needs to send the ID of the current round in a short beacon (see \cref{fig:ttw_overview}).
The nodes can then use the ID to look up the corresponding scheduling table in their local memory.
As a result, the rounds in \TTnet are ``more compact,'' which minimizes delays and improves efficiency.

The beacons also contain all information required to reliably switch between different operation modes at runtime.
In addition to the round ID, each beacon carries a mode ID and a trigger bit.
Modes and rounds have unique IDs with a known mapping of rounds to modes.
By default, a beacon includes the current mode ID and the trigger bit is set to $0$.
A mode change happens in two phases.
First, the mode change is announced by a beacon with the new mode ID.
In a later round, the trigger bit is set to $1$, which triggers the mode change; the new mode starts at the end of that round.
This two-step procedure lets the nodes prepare for an upcoming mode change; for example, by stopping the execution of application tasks that are not present in the new mode.
Hence, \TTnet provides a generic interface to implement different mode-change protocols.
The mode-change protocol itself is independent of \TTnet and can be freely designed or chosen among existing protocols~\cite{chen2018SafeMC}.

Sending beacons in every round ensures a safe operation of \TTnet:
If a node fails to receive one, it does not participate in the communication until it receives again a beacon.
Thus, it is guaranteed that all participating nodes know the current round and mode IDs.%
}

\subsection{\TTnet Implementation}
\label{subsec:ttnet_implem}

We implement \TTnet using \baloo~\cite{jacob2019Baloo}, a design framework for communication stacks based on synchronous transmissions. \baloo provides a customizable round structure and allows the user to implement the high-level logic of a communication stack via a programming interface based on callback functions.
Low-level operations such as time synchronization and radio control are handled by a middleware layer integrated in \baloo.

\squarepar{%
Our \TTnet implementation uses the static configuration mode of \baloo. The beacons are implemented using the user-bytes field of \baloo's control packets with a payload size of 2 bytes. The implementation runs on any physical platform supported by \baloo. Because of the need for timing predictability in \TTW, we focus on a platform that is based on the dual-processor platform (\DPP) architecture~\cite{beutel2019DPPdemo}.
The \DPP combines two arbitrary processors with an interconnect called \bolt~\cite{sutton2015Bolt}. \bolt provides predictable asynchronous message passing between two processors using message queues with first-in-first-out (FIFO) semantics, one for each direction. This allows to dedicate each processor to a specific task, execute these tasks in parallel, and compared to traditional dual- or multi-core platforms the \DPP offers formally verified bounds on the interference between the processors~\cite{sutton2015Bolt}.
The specific platform we use is composed of a 32-bit ARM Cortex-M4 based MSP432P401R running at 48\MHz and a 16-bit CC430F5147 running at 13\MHz. The former is dedicated to the execution of application tasks (\eg sensing, actuation, and control), while the latter is dedicated to \TTnet using a low-power wireless radio operating at 250\kbps in the 868\MHz frequency band.
\final{Our implementation is open source and freely available; refer to~\cite{repoTTW} for details.}%
}

\subsection{\TTnet Model}
\label{subsec:ttnet_model}

The real-time scheduler in \TTW requires timing and energy models of \TTnet's operation to synthesize the scheduling tables.
These models must be \final{tight bounds to avoid delays and energy costs.}

\begin{figure}
\includegraphics[scale=0.8]{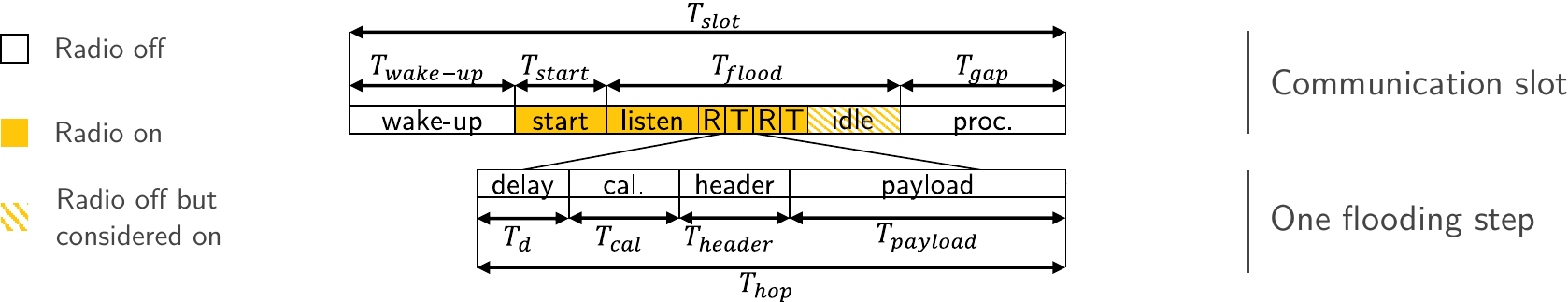}
\caption{%
Detailed timings during a slot in \TTnet.
\capt{%
At the slot level, the colored boxes identify phases where the radio is on.
In the idle phase, the radio is off, but the duration depends on a node's distance to the initiator of the Glossy flood.
To estimate the energy savings of rounds, we make the pessimistic assumption that the radio is on for \Tglossy, as specified in~\cref{eq:Ton}.
}
}
\label{fig:Tslot}
\end{figure}

Let \Tround be the length of a \TTnet round.
A round consists of a beacon slot plus up to \nslotsmax regular slots, in which Glossy floods are executed (see \cref{fig:ttw_overview}).
$\Tround(L,\nslots)$ denotes the length of a round with \nslots regular slots having the same payload size $L$.
The structure of a slot is illustrated in~\cref{fig:Tslot}.
Accordingly, the length of a slot \Tslot can be decomposed as follows
\begin{equation}
  \Tslot = \Twakeup + \Tstart + \Tglossy + \Tgap
\end{equation}
First, the nodes wake up (\Twakeup) and switch on their radio (\Tstart).
Then, the Glossy flood executes (\Tglossy) followed by a small gap time (\Tgap) in which the nodes can process the received packet.
\final{As explained in~\cite{zimmerling2015dissertation},} the total length of a flood \Tglossy can be expressed by
\begin{align}
\Tglossy = (H+2N-1)*\Thop
\end{align}
where $H$ is the network diameter and $N$ is the \final{maximum number of times a node transmits} during a Glossy flood. Let \Thop be the time needed for one protocol step, that is, the duration of one (synchronous) transmission during a flood.
This quantity can be decomposed as%
\begin{align}
\Thop = \Td + \Tcal + \Theader + \Tpayload
\end{align}
where \Td is an initial radio delay, \Tcal is the time needed to calibrate the radio clock \final{(taking time equal to the transmission of \Lcal bytes), and \Theader and \Tpayload are the times needed to transmit the packet header consisting of \Lheader bytes and the message payload consisting of $L$ bytes}.
Using a radio with a transmit bitrate of \Rbit, the transmission of $L$\bytes takes
\begin{align}
T(L) = 8L/\Rbit
\end{align}
We divide the length of a slot \Tslot into \Ton and \Toff, the times spent with the radio on and off, respectively.
To this end, we make the conservative assumption that the radio stays on for the entirety of \Tglossy, as illustrated in \cref{fig:Tslot} and explained in the caption.
\begin{align}
\Tslot(L) &\,=\,
  \Toff + \Ton(L) \\
\Ton(L) &\,=\,
	\Tstart + (H+2N-1)*\left( \Td + 8(\Lcal + \Lheader + L)/\Rbit \right) \\
\Toff &\,=\,
	\Twakeup + \Tgap
\label{eq:Ton}
\end{align}
With this, the length of a round with \nslots regular slots and a common payload size $L$ is
\begin{equation}
 \Tround(L, \nslots) = \Tslot(\Lbeacon) + \nslots*\Tslot(L) + \Tpreprocess
\end{equation}
where \Lbeacon is the payload size of a beacon and \Tpreprocess is the time needed by our \TTnet implementation to prepare for an upcoming round (\eg retrieve messages from send queue).

In addition to the timings of a \TTnet round, we derive a model for the relative energy savings obtained by grouping multiple slots into the same round.
As discussed in \cref{subsec:ttnet_design}, sending beacons is necessary to ensure a safe \final{protocol} operation.
Without rounds, each regular message must still be preceded by a beacon to provide the same guarantee.
In this case, the transmission time \Tworound for \nslots regular messages of size $L$ is given by
\begin{align}
\Tworound(L,\nslots) =  \nslots*( \, \Tslot(\Lbeacon) + \Tslot(L) \, )
\end{align}
The relative energy savings obtained by a round-based design thus \final{amount} to
\begin{equation}
E= (\Tworoundon - \Troundon)/\Tworoundon
\end{equation}
\final{with $\Tworoundon = \nslots*( \, \Ton(\Lbeacon) + \Ton(L) \, )$ and $\Troundon = \Ton(\Lbeacon) + \nslots * \Ton(L)$.}

\begin{figure}
  \href{\ttwfig{Figure-3}}{%
  \includegraphics[scale=0.8]{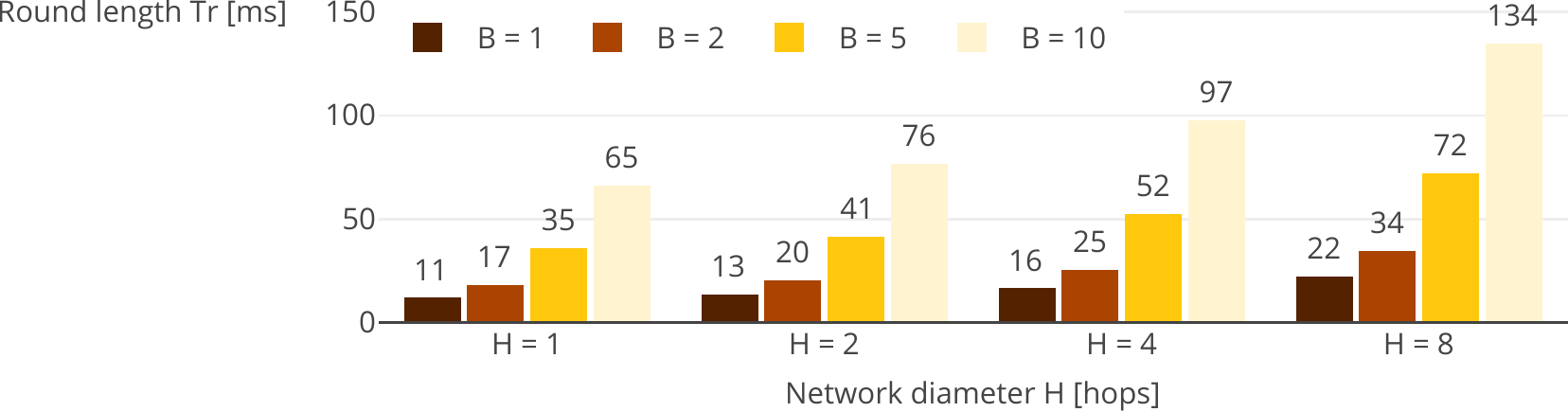}}
  \caption{Example values of round length computed using the \TTnet model for a payload size of 16\bytes and $N = 2$ transmissions during a Glossy flood.
	\capt{%
	Fewer slots lead to shorter rounds (\eg 52\ms for $\nslots=5$ slots across a 4-hop network), which ultimately allows for shorter end-to-end delays.
  }}
  \label{fig:TTWmodel}
\end{figure}

Using these expressions and the parameters measured for or used by our \TTnet implementation (detailed in \cite{repoTTW}), we can determine the round length \Tround and the energy savings $E$ for different number of slots per round \nslots, message payload sizes $L$, network diameters $H$, and number of transmissions during a Glossy flood ($N$).
For $L=16$ byte and $N=2$, \cref{fig:TTWmodel} plots the resulting round length \Tround for different network diameters and slots per round.
We can see, for example, that it takes less than 100\ms to send 10 messages in a 4-hop network.
\final{Real-world experiments using 27 distributed wireless nodes on a public testbed demonstrate that these models are very accurate~(\cref{subsec:ttnet_eval}).}


\section{Real-Time Scheduler}
\label{sec:scheduler}

\squarepar{%
Equipped with a timing-predictable implementation of the \TTnet communication stack and a corresponding model that provides accurate upper bounds, we now turn to the second main component of \TTW: the real-time scheduler.
This section first defines the system model and scheduling problem, then describes the single-mode and multi-mode schedule synthesis.%
}

\subsection{System Model}
\label{subsec:systModel}

\squarepar{%
Before we can formulate and solve the scheduling problem we face in \TTW, we provide a formal system model of all relevant hardware and software components and their interactions.

\fakepar{Platform}
Each node in the system is a \emph{platform} that executes tasks and exchanges messages using \TTnet.
All tasks have a known worst-case execution time (WCET) on that platform.
Targeting state-of-the-art wireless \cps platforms with two processors (\eg LPC541XX, VF3xxR, \DPP~\cite{beutel2019DPPdemo}), each node can simultaneously execute tasks and exchange messages.%
}

\begin{figure}[!tb]
\centering
\includegraphics[width=0.7\linewidth]{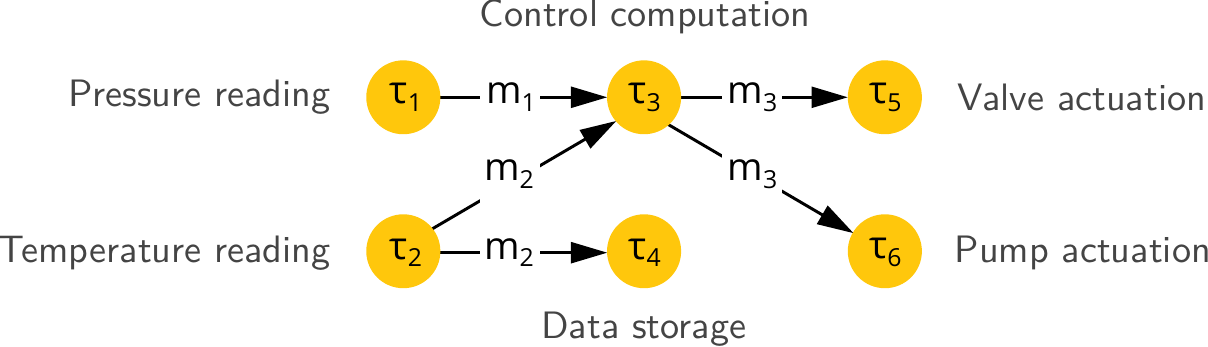}
\vspace{-1mm}
\caption{An example application and its precedence graph \predG.
\capt{The application execution starts with sensing tasks $\tau_1$ and $\tau_2$. After messages $m_1$ and $m_2$ with the sensor readings are received by the controller, actuation values are computed ($\tau_3$), sent to the actuators ($m_3$), and applied ($\tau_5$ and $\tau_6$).
}}
\vspace{-3mm}
\label{fig:precedence_graph}
\end{figure}

\fakepar{Applications}
Let \appset denote the set of distributed \emph{applications} in the system.
An application is composed of \emph{tasks} and \emph{messages}, each with a unique task or message \id, that are coupled by precedence constraints as described by a directed acyclic graph, where vertices and edges represent tasks and messages, respectively. We denote by \app.\predG the \emph{precedence graph} of application \app~(see \cref{fig:precedence_graph} for an example).
Each application executes at a periodic interval $\app.p$ called the \emph{period}.
An application execution is completed when all tasks in \predG have been executed.
All tasks and messages in \app.\predG share the same period $\app.p$.
\final{If the same task belongs to applications with different periods, it can be equivalently modeled as two different tasks.}
Applications are subject to real-time constraints:
an application \emph{relative deadline}, denoted by $\app.d$, represents the maximum tolerable \emph{end-to-end delay} to complete the application execution.
The deadline can be arbitrary, without a specific relation to the period $\app.p$.
Some applications may require to keep the same schedule (\eg the same task offsets) when switching between different operation modes, for example, to guarantee the stability of control loops under arbitrary mode switches~\cite{baumann2019TCPS}.
We call these \emph{persistent applications} and denote them by $\persappset \subset \appset$.

\fakepar{Tasks}
We denote by \taskset the set of all \emph{tasks} in the system.
A node executes at most one task at any point in time.
Since interactions with the physical world like sensing and actuation should not be interrupted, we consider non-preemptive task scheduling.
Each task $\tau$ is mapped to a given node $\tau.map$ on which it executes with WCET $\tau.e$.
The \emph{task offset} $\tau.o$ represents the start of the task execution relative to the beginning of the application execution.
A task can have an arbitrary number of preceding messages, which must all be received before the task can start to execute.
$\tau.prec$ denotes the set of preceding message \ids.
Within one application, each task is unique; however, the same task may belong to multiple applications (\eg the same sensing task may source different feedback loops). If so, we consider that these applications have the same period.

\fakepar{Messages}
Let \messageset be the set of all \emph{messages}.
Every message $m$ has at least one preceding task, that is, a task that needs to finish before the message can be transmitted. The set of preceding task \ids is denoted by $m.prec$.
The \emph{message offset} $m.o$ relative to the beginning of the application execution represents the earliest time message $m$ can be allocated to a \TTnet round for transmission (\ie after all preceding tasks are completed).
The \emph{message deadline} $m.d$ relative to the message offset represents the latest time the message transmission must be completed (\ie the earliest offset of subsequent tasks).
The payload of all messages is upper-bounded by \Lmax.
Messages are not necessarily unique: multiple edges of \app.\predG can be labeled with the same message $m$, which captures the case of multicast or broadcast transmissions~(see \cref{fig:precedence_graph}).
If a message belongs to multiple applications, we consider that these applications have the same period.

\fakepar{Modes}
We denote with \modeset the set of operation \emph{modes}.
These modes represent mutually exclusive phases in the system execution (\eg init, normal, and failure modes), each having its own schedule.
A mode has a unique \emph{priority} \modeany.\prio. 
We write $\app \in \modeany$ to denote that application \app executes in mode \modeany.
When unambiguous, we use \modeany to denote the set of all applications that execute in mode \modeany.
The \emph{hyperperiod} \modeHyperperiod of mode \modeany is the least common multiple of the periods of all applications in \modeany.
Possible transitions between modes at runtime are specified by the \emph{mode graph} \modeGraph~(see \cref{fig:modeGraph}). The mode graph is undirectional, that is, a transition from \modei to \modej implies that it is also possible to switch from \modej to \modei.


\fakepar{Network and rounds}
We denote with \nodeset the set of all \emph{nodes} in the network.
The following four network parameters must be specified by the user of the \TTW architecture:
\begin{itemize}
	\item \customBox{$L$} the payload size of regular messages, in bytes;
	\item \customBox{\nslotsmax} the maximum number of slots in a \TTnet round;
	\item \customBox{$N$} the maximum number of transmissions of a node during a Glossy flood;
	\item \customBox{$H$} the estimated diameter of the network, in number of hops.
\end{itemize}
As explained in \cref{sec:ttnet}, communication over the network is scheduled in \emph{rounds} $r$.
The schedule of mode \mode{} has $R_{\mode{}}$ rounds.
We consider \TTnet rounds as atomic, that is, they cannot be interrupted. Thus, the ordering of messages in a round is irrelevant.

Round $r$ contains \nslotsround slots (at most \nslotsmax), each of which is allocated to a unique message $m$.
The \emph{allocation vector} $r.[\nslotsround]$ contains the \ids of the messages that are allocated to the slots in round $r$, where $r.B_s$ denotes the allocation of the $s$-th slot.
The \emph{starting time} $r.t$ is the start of the round relative to the beginning of the mode's hyperperiod.
Using the models from \cref{subsec:ttnet_model}, the length of a round and its energy cost can be determined.

\begin{table}[!tb]
	\caption{Inputs and outputs of the scheduling problem we solve in \TTW}
	\vspace{-3mm}
	\label{table:ttw_inputs_outputs}
	\smaller{\begin{tabularx}{\linewidth}{ @{}l@{\qquad}l@{\qquad}X@{}}
\toprule
\multirow{11}{*}{\textbf{Inputs}}&$\nodeset$&Set of nodes in the system\\
&$\multirow{2}{*}{\appset,\, \persappset}$&Set of applications and persistent applications\\
&&(including periods, deadlines, and precedence graphs)\\
&$\modeset$&Set of operation modes (including mode priorities)\\
&$\modeGraph$&Mode graph\\
&$\tau.p,\, m.p$&Task and message periods, inherited from the application\\
&$\tau.\map$&Mappings of tasks to nodes\\
&$\tau.e$&Task worst-case execution time (WCET) given $\tau.\map$\\

&$\nslotsmax$&Maximum number of slots per round\\
&$\Lmax$&Maximum message payload size\\
&$H$&Network diameter (in number of hops)\\
&$N$&Number of transmissions in a Glossy flood~\cite{ferrari2011Glossy}\\
\midrule
\multirow{3}{*}{\textbf{Outputs}}&$\tau.o$&Task offsets\\
&$m.o,\, m.d$&Message offsets and deadlines\\
&$r.t,\, r.[\nslotsmax]$&Round starting times and allocation vectors\\
\bottomrule
\end{tabularx}
}
	\vspace{-3mm}
\end{table}

\subsection{Scheduling Problem}
\label{subsec:problem}

Based on our system model, we are now in the position to formally state the scheduling problem we have to solve in \TTW.
Given all modes, applications, task-to-node mappings, and WCETs, for each mode \mode{} the remaining variables define the \emph{mode schedule} $\sched{\mode{}}$
\begin{align*}
&\sched{\mode{}} \, = \,
	\left\lbrace
	\begin{tabular}{c|l}
	$\tau.o, \, m.o, \, m.d$
	&
	$\forall \; \app \in \mode{}, \;
	(\tau,m) \in \app.\predG$
	\\
	$r_k.t, \, r_k.[\nslotsmax]$
	&
	$\forall \; k \in [1, \, R_{\mode{}}]$
	\end{tabular}
	\right\rbrace
\end{align*}
\cref{table:ttw_inputs_outputs} lists all inputs and outputs of the scheduling problem in \TTW.
A schedule for mode \modeany is said to be \emph{valid} if all applications executing in mode \modeany meet their end-to-end deadlines.
The scheduling problem to solve thus consists of finding valid schedules for all modes $\modeany \in \modeset$ such that the following two objectives are met:
\begin{description}
	\item [\objective{1}]
	The number of communication rounds is minimized, which results in minimizing the energy required for wireless communication.
	\item [\objective{2}]
	All persistent applications $\persappset \subset \appset$ seamlessly switch between modes, that is, their schedules remain the same across all possible mode changes.
\end{description}


\subsection{Single-Mode Schedule Synthesis}
\label{sec:single_mode}

\squarepar{%
\TTW statically synthesizes the schedule of all tasks, messages, and communication rounds by solving a MILP.
We first look at the single-mode case, effectively showing how to achieve objective \objective{1}, whereas \cref{sec:multi_mode} considers the multi-mode case and thus objective \objective{2}.

The schedule of a mode \modeany is computed for one hyperperiod, after which it repeats itself.
To minimize the number of rounds used while taming computational complexity, we solve the problem sequentially as described in~\cref{alg:outerlayer}.
Each MILP formulation considers a fixed number of rounds $R_{\mode{}}$ to be scheduled, starting with $R_{\mode{}}=0$. The number of rounds is incremented until a feasible solution is found or the maximum number of rounds $R_{max}$ that fit into one hyperperiod is reached.
So, even without an explicit objective function, \cref{alg:outerlayer} guarantees by construction that if the problem is feasible, the synthesized schedule uses the minimum number of rounds, thereby \final{minimizing} the communication energy costs.%
}

\begin{algorithm}[!tb]
\begin{algorithmic}
\smaller
\Require
	mode \mode{},\,
	$\app \in \mode{}$,\,
	$\tau.\map$,\,
	$\tau.e$,\,
	\nslotsmax,\,
	\Toffset
\Ensure
	\sched{M}

\State $LCM$ = \textit{hyperperiod}(\mode{})
\State $R_{max} = floor(LCM/\Toffset)$
\State $R_{\mode{}} = 0$

\While{$R_{\mode{}} \leq R_{max}$}
	\State formulate the MILP for mode \mode{} using $R_{\mode{}}$ rounds
	\State ( \sched{M} , \textit{feasible} ) = \textit{solve}(MILP)
	\If {\textit{feasible}}
		\Return \sched{M}
	\EndIf
	\State $R_{\mode{}} = R_{\mode{}}+1$
\EndWhile
\State \Return 'Problem infeasible'
\end{algorithmic}
\caption{\small Pseudo-code of the single-mode schedule synthesis}
\label{alg:outerlayer}
\end{algorithm}

Each MILP formulation contains a set of classical scheduling constraints such as: precedence constraints between tasks and messages must be respected, the applications' end-to-end deadlines must be satisfied, nodes process at most one task simultaneously, communication rounds must not overlap, and rounds must not be allocated more then \nslotsmax messages.
These constraints can be easily formulated based on our system model. 
However, we must also guarantee that the allocation of messages to rounds is valid. Specifically,
\begin{description}
	\item [\constraint{1}]
	messages must be served in rounds that start after their release time;
	\item [\constraint{2}]
	messages must be served in rounds that finish before their deadline.
\end{description}
In other words, we must integrate the bin-packing problem that arises when allocating messages to rounds within the MILP formulation.
This is a non-trivial challenge and a major difference compared with the existing approaches for wired architectures~(\eg \cite{craciunas2016Combined}).

\begin{figure}
\includegraphics[scale=0.8]{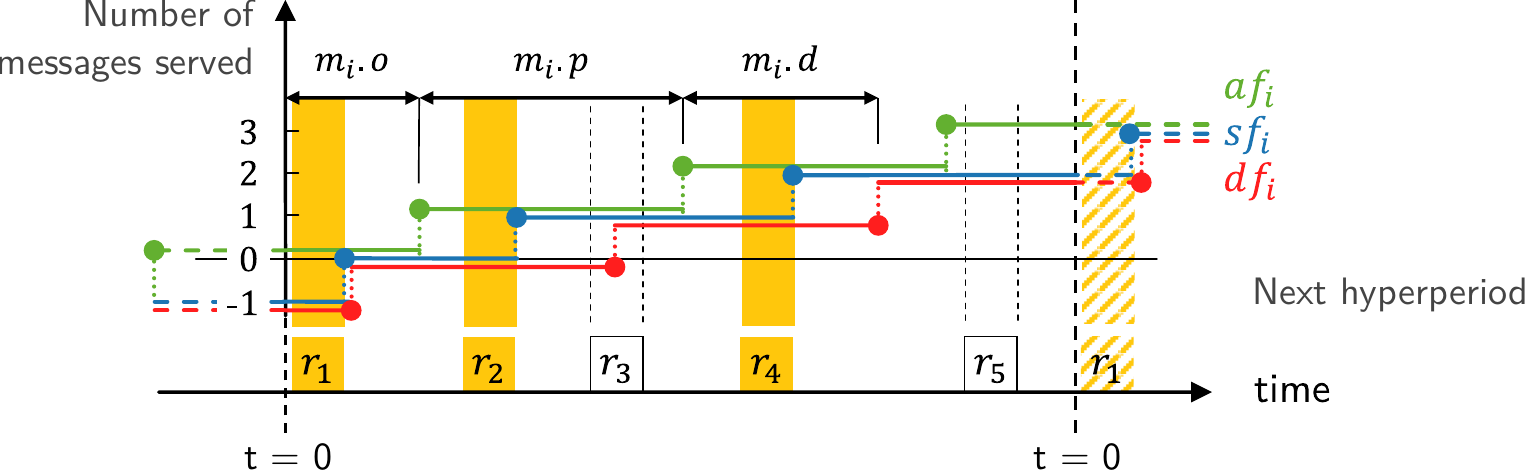}
\caption{Example showing arrival (\af), demand (\df), and service (\sf) functions for message $m_i$.
\capt{%
The lower part of the chart shows the five round, $r_1$ to $r_5$, that are scheduled within the hyperperiod.
Message $m_i$ is allocated a slot in the colored rounds, that is, in rounds $r_1$, $r_2$, and $r_4$.
Allocating $m_i$ to $r_3$ instead of $r_2$ would be invalid because $r_3$ does not finish before $m_i$'s deadline, thus violating \final{constraint} \constraint{2}.
By contrast, allocating $m_i$ to $r_5$ instead of $r_1$ would be valid and result in $r_0.B_i = 0$.}
}
\vspace{-2mm}
\label{fig:afdfsf}
\end{figure}

To address this challenge, we first formulate the constraints \constraint{1} and \constraint{2} using \emph{arrival}, \emph{demand}, and \emph{service} functions---denoted by \af, \df, and \sf---inspired by network calculus~\cite{leboudec2001Network} (see \cref{fig:afdfsf} for an illustration).
These three functions count the number of message instances that are released, have passed deadlines, and have been served since the beginning of the hyperperiod.
It must hold that
\begin{flalign}
\label{eq:df<sf<af}
&\forall\, m_i \in \messageset, \;\forall\, t,
&&\df_i(t) \leq \sf_i(t) \leq \af_i(t)
&&\\
\label{eq:af_def}
&\text{with}
&&\af_i: \; t \;
	\longmapsto \; \left \lfloor{\frac{t-m_i.o}{m_i.p}}\right \rfloor 	+ 1
	&&\\
\label{eq:df_def}
&\text{and}
&&\df_i: \; t \;
	\longmapsto \; \left \lceil{\frac{t-m_i.o-m_i.d}{m_i.p}}\right \rceil
	&&
\end{flalign}

\noindent
However, as the service function \sf stays constant between the rounds, we can formulate constraints \constraint{1} and \constraint{2} as follows:
$\forall\, m_i \in \messageset, \; \forall\, j \in [1 .. R_{\mode{}}], $
\begin{flalign}
\label{eq:af_const}
&\textup{\constraint{1}} \quad   \quad
	&\sf_i(r_j.t + \Tround) \, &\leq \, \af_i(r_j.t)
	&&
\\
\label{eq:df_const}
&\textup{\constraint{2}} \quad   \quad
	&\sf_i(r_j.t)  \, &\geq \, \df_i(r_j.t + \Tround)
	&&
\end{flalign}

\noindent
The arrival and demand functions are step functions, which cannot be directly used in a MILP formulation. However, we observe that
\begin{flalign}
\label{eq:af=k}
&\forall \; k \in \mathbb{N}, \quad
&&\af_i(t) = k
	\quad \Leftrightarrow \quad
	0 \, \leq \, t - m_i.o - (k-1)*m_i.p \,<\, m_i.p &&\\
&\text{and} 
&&\df_i(t) = k
\label{eq:df=k}
	\quad \Leftrightarrow \quad
	0 \, < \, t - m_i.o - m_i.d - (k-1)*m_i.p \,\leq\, m_i.p &&
\end{flalign}
This allows us to introduce, for each message $m_i\in \messageset$ and each round $r_j$, $j \in [1..R_{\mode{}}]$, two integer variables $k^a_{ij}$ and $k^d_{ij}$ that we constrain to take the values of \af and \df at the time points of interest, namely $r_j.t$ and $r_j.t + \Troundj$. More formally, we can write
\begin{align}
\label{eq:ka} 
0 \, \leq \, r_j.t
	&-m_i.o - (k^a_{ij}-1)*m_i.p \,<\, m_i.p\\
\label{eq:kd} 
0 \, < \, r_j.t
	&+\Troundj - m_i.o - m_i.d - (k^d_{ij}-1)*m_i.p \,\leq\, m_i.p\\
\notag
\text{Thus,} \hspace{15pt} &\eqref{eq:ka} \quad \Leftrightarrow  \quad
	 \af_i(r_j.t) = k^a_{ij} \\
\notag
	&\eqref{eq:kd} \quad \Leftrightarrow \quad
	\df_i(r_j.t + \Troundj) = k^d_{ij}
\end{align}

\noindent
Finally, we must express the service function \sf, which counts the number of message instances served by \emph{the end} of each round.
Recalling that $r_k.B_s$ denotes the \id of the message that is allocated to the $s$-{th} slot of round $r_k$, we have that for any time \mbox{$t \in \; [ \; r_{j}.t + \Troundj \, ; \,  r_{j+1}.t + \Troundj \; [$} the number of instances of message $m_i$ served is
\begin{align*}
	\sum_{\substack{k = 1}}^{j} \;\;
	\sum_{\substack{s = 1}}^{B}
	 \; r_k.B_s
	 \quad s.t. \quad B_s = i
\end{align*}

\noindent
As visible in \cref{fig:afdfsf}, it can happen that $m.o + m.d > m.p$, resulting in \mbox{$\df(0)=-1$} according to \cref{eq:df_def}.
This situation arises when a message is released at the end of a hyperperiod and thus has its deadline in the \emph{next} hyperperiod.
To account for this, we introduce for each message $m_i$ a variable $r_0.B_i$ that is set to the number of such ``leftover'' message instances at $t=0$. With this, for each $m_i \in \messageset$ and $t \in \; [ \; r_{j}.t + \Troundj \, ; \,  r_{j+1}.t + \Troundj \; [$,
\begin{flalign}
\label{eq:sf_def}
\sf_i: \; t \;
	&\longmapsto \;
	\sum_{\substack{k = 1 \\[2pt]s.t. \; r_k.t + \Troundk \, < \, t}}^{j}
	\quad
	\sum_{\substack{s = 1 \\[2pt]s.t. \; B_s = i}}^{B}
	 r_k.B_s - r_0.B_i
\end{flalign}

\noindent
Based on the above reasoning, we can express constraints \constraint{1} and \constraint{2} in the MILP formulation using \cref{eq:ka,eq:kd} and the following two equations
\begin{align}
&
\eqref{eq:af_const}\quad	 \Leftrightarrow
	\quad
	\sum_{k = 1}^j
	\sum_{\substack{s = 1 \\[2pt]s.t. \; B_s = i}}^{B}
	 \; r_k.B_s - r_0.B_i \; \leq\;  k^a_{ij}
\\
&
\eqref{eq:df_const}\quad	 \Leftrightarrow
	\quad
	\sum_{k = 1}^{j-1}
	\sum_{\substack{s = 1 \\[2pt]s.t. \; B_s = i}}^{B} \; r_k.B_s - r_0.B_i
	\; \geq\;
	k^d_{ij}
\end{align}


\subsection{Multi-Mode Schedule Synthesis}
\label{sec:multi_mode}

The multi-mode schedule synthesis is a multi-objective problem:
As stated in \cref{subsec:problem}, the number of communication rounds in each mode should be minimized \objective{1}, while the schedule of all persistent applications should remain unchanged across mode changes \objective{2}.
Objective \objective{2} induces dependencies between the different modes, which cannot be handled by solving a set of independent single-mode schedule synthesis problems.

\squarepar{%
A straightforward approach would be to synthesize the schedules of all modes at once based on a single MILP formulation.
This approach, however, has two caveats.
First, the resulting schedule synthesis problem is NP-hard~\cite{jeffay1991nonpreemptive} and thus scales poorly as the number of modes increases, which becomes a bottleneck for realistic \CPS applications requiring a high degree of system adaptability.
Second, a global objective function must be defined, such as minimizing the total number of rounds across all modes, which still gives no guarantee that the corresponding communication energy costs are effectively minimized: at runtime, if the system remains almost always in a certain mode, it may be better to minimize the number of rounds in that particular mode even if this implies a larger number of rounds overall.%
}

\begin{algorithm}[!tb]
\begin{algorithmic}
\smaller
\Require
	Applications specification; modes specification; network parameters; system parameters
\Ensure
	\{\, \sched{\modei} for \modei $\in$ \modeset \}
\vskip 2pt
\State \textit{inheritance\_constraints} = $\emptyset$

\For{all \modei $\in$ \modeset in order of decreasing mode priority \modei.\prio}
	\State add \textit{inheritance\_constraints} to mode \modei
	\State ( \sched{\modei} , \textit{feasible} ) = \textit{single\_mode\_synthesis}(\modei)
	\If {\textit{feasible}}
		\State add \sched{\modei} to \textit{inheritance\_constraints}
	\Else
		\State \Return 'Problem infeasible'
	\EndIf
\EndFor
\State \Return \{\, \sched{\modei} for \modei $\in$ \modeset\}
\end{algorithmic}
\caption{\small Pseudo-code of the multi-mode schedule synthesis}
\label{alg:multi_mode}
\end{algorithm}

To address these caveats, we solve the multi-mode schedule synthesis problem sequentially using heuristics, as commonly done in related approaches~\cite{joshi2017MultiDomain,pozo2015SMTbased,steiner2010evaluation}.
\cref{alg:multi_mode} summarizes our approach.
The modes are scheduled based on their priority \modeany.\prio.
The mode with the highest priority (\ie $\modeany.\prio=1$) is scheduled first according to the single-mode synthesis outlined in \cref{alg:outerlayer}.
Then, the mode with the second-highest priority (\ie $\modeany.\prio=2$) is considered; however, its specification is first extended with \emph{inheritance constraints} to guarantee that all persistent applications can seamlessly switch between all modes considered so far.
The process repeats until all modes are scheduled.

This sequential approach addresses the scalability of the multi-mode problem and provides a reasonable heuristic for minimizing the number of executed rounds and thus the communication energy costs.
Indeed, these costs ultimately depend on how much time the system spends in each mode.
It is reasonable to assume that an application domain expert knows in which modes the system likely operates most of the time.
These modes are assigned a higher priority and therefore scheduled first by \cref{alg:multi_mode}.
A mode with a lower priority is subject to more inheritance constraints, so it may schedule more than the minimal number of rounds to achieve \objective{2}.
This may lead to a sub-optimal energy cost of lower-priority modes, which is nevertheless acceptable as the system likely spends less time in these modes.

In the following, we address the problem of deriving the set of inheritance constraints that are \emph{necessary} and \emph{sufficient} to guarantee objective~\objB.
We first formalize the continuity constraints necessary to satisfy~\objB and characterize how these constraints may lead to conflicts between the mode schedules.
Then we derive the minimally restrictive inheritance constraints that are necessary and sufficient to prevent such conflicts while satisfying \objB.

\begin{figure}
\centering
\includegraphics[width=0.5\linewidth]{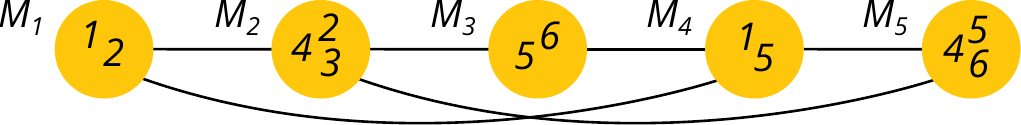}
\caption{%
Mode graph \modeGraph for \cref{exp:sched_conflict_basic,exp:sched_domain}.
\capt{Five modes are depictd by circles and possible transitions between them by edges. Numbers inside each circle indicate which of the six applications, \appl{1} to \appl{6}, execute in the corresponding mode, for example, $M_1 = \{\appl{1},\appl{2}\}$. Mode \modei has priority $i$.}}
\label{fig:modeGraph}
\vspace{-1mm}
\end{figure}

\begin{figure}
	\begin{subfigure}[t]{.48\linewidth}
		\centering
		\includegraphics[scale=0.9]{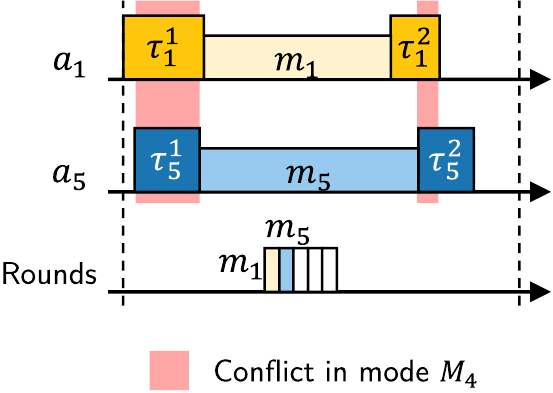}
		\caption{%
		\appl{5} is scheduled in mode \mode{3} without considering the previously computed schedule of \appl{1}, which leads to a conflict in mode \mode{4}.}
		\label{subfig:example_sched_conflict}
	\end{subfigure}%
	\hfill
	\begin{subfigure}[t]{.48\linewidth}
		\centering
		\includegraphics[scale=0.9]{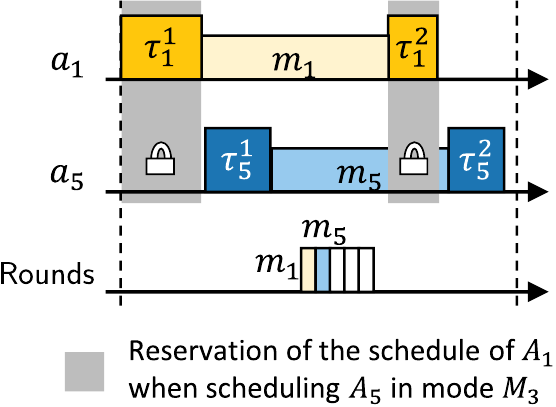}
		\caption{%
		\appl{5} is scheduled in mode \mode{3} considering the schedule of \appl{1} as reserved. A compatible schedule for \appl{5} is found, preventing a conflict in \mode{4}.}
		\label{subfig:example_sched_reservation}
	\end{subfigure}
	\vspace{-1mm}
	\caption{%
		Representations of the schedule of applications \appl{1} and \appl{5} from~\cref{exp:sched_conflict_basic}.
		For sake of illustration, we consider that all tasks are mapped to the same node.
		\appl{1} and \appl{5} are scheduled in modes \mode{1} and \mode{3}, and must both be inherited in mode \mode{4}.
		In~\cref{subfig:example_sched_conflict}, overlapping task schedules result in a conflict, while in~\cref{subfig:example_sched_reservation} this was prevented by reserving \appl{1}'s schedule.
	}
	\vspace{-4mm}
	\label{fig:example_sched}
\end{figure}

\vspace{-1.5mm}
\begin{example}
\squarepar{%
\label{exp:sched_conflict_basic}
	Let us consider the mode graph in \cref{fig:modeGraph} and assume that all applications are persistent.
	The modes are scheduled sequentially, starting with the highest-priority mode \mode{1}, which is freely scheduled.
	When mode \mode{2} is scheduled, the schedule for application \appl{2} is inherited from mode \mode{1} to guarantee \objective{2} and the schedules for applications \appl{3} and \appl{4} are synthesized without constraints.
	In \mode{3}, the specified applications \appl{5} and \appl{6} are new and can be scheduled without constraints.
	Then, in mode \mode{4}, the specified applications \appl{1} and \appl{5} have been previously scheduled and thus must be inherited~\objective{2}.
	However, as mode \mode{3} has been scheduled without constraint, the schedule synthesized for \appl{5} may be incompatible with that of \appl{1} from mode \mode{1}. This leads to a conflict in \mode{4} and thus renders the sequential synthesis of the multi-mode problem infeasible, as illustrated in \cref{fig:example_sched}.}
\end{example}

\vspace{-4mm}
\subsubsection{Continuity Constraints}
\label{subsec:continuity_constraints}
\vspace{-1mm}

\squarepar{%
The schedule synthesis returns the application schedules (\ie task and message offsets, and message deadlines) and the round schedules (\ie starting times and allocation vectors).
We represent an application schedule through a scheduling function $s$ as defined below.%
}

\begin{definition}[Scheduling function]
\label{def:sched_funtion}
The scheduling function $s$, defined over the set of applications \appset, returns for a given application \app all parameters characterizing the schedule of \app. The schedule of application \app is denoted by $s(\app)$. $s_\mode{}(\app)$ denotes the schedule of application \app in mode \modeany.
The scheduling function is extended to sets of applications as follows
\begin{equation*}
\forall \, S \subset \appset \; , \quad s(S) = \bigcup_{\app\in S} s(\app)
\end{equation*}

\end{definition}

All persistent applications $\app \in \persappset$ must keep the same schedule across mode changes, which leads us to the definition of continuity constraints.
\begin{definition}[Continuity constraint]
\label{def:continuity_constraint}
The set of all continuity constraints is given by
\begin{equation}
\forall\, \app \in \persappset, \, \forall\, (\modei, \modej) \in \modeset^2, \, \app \in \modei \, \wedge \, \app \in \modej \ \wedge \, \modeGraph(\modei,\modej) = 1
	\; \Rightarrow \;
	s_\modei(\app) = s_\modej(\app)
\end{equation}
\end{definition}

\begin{definition}[Schedule domains]
\label{def:sched_domains}
The schedule domains of an application are the possibly multiple subsets of modes in which the application schedule must remain the same.
\end{definition}

\begin{corollary}
	Two modes \modei and \modej belong to the same schedule domain of an application $\app \in \persappset$ if and only if
    (i) \app is scheduled in both modes, that is, $\app \in \modei \; \wedge \; \app \in \modej$, and (ii) there is a possible transition between the two modes, that is, $\modeGraph(\modei, \modej) = 1$.
\end{corollary}

\proof
Multiple modes belong to the same schedule domain because of a continuity constraint.
The formalization of the schedule domains directly follows from~\cref{def:continuity_constraint}
\qed

\squarepar{%
The schedule domains of an application can be \final{extracted} from the mode graph \modeGraph.
One approach entails considering the sub-graph \modeGraphA of \modeGraph that contains only the modes in which application \app is specified---every connected component of \modeGraphA is a schedule domain of \app.

Any non-persistent application that is present in multiple modes can be replaced by a distinct application with the same parameters in the respective modes.
Similarly, any persistent application with multiple schedule domains can be replicated into distinct applications having one scheduling domain each (illustrated by \cref{exp:sched_domain}).
This leads us to consider in the following that all applications are persistent and have a single schedule domain.%
}

\begin{example}\label{exp:sched_domain}
\squarepar{%
Consider again the mode graph in \cref{fig:modeGraph}. Application \appl{6} has two schedule domains, $\{\mode{3}\}$ and $\{\mode{5}\}$. This can also be modeled by two distinct applications \appl{6.3} and \appl{6.6} executing in \mode{3} and \mode{6}. Application \appl{1} has only one schedule domain, $\{\mode{1},\mode{4}\}$.
If \appl{1} was not persistent, the continuity constraint would not apply and \appl{1} could be equivalently modeled by two distinct applications \appl{1.1} and \appl{1.4} executing in \mode{1} and \mode{4}.%
}
\end{example}

Continuity constraints can cause conflicts leading to the failure of the multi-mode synthesis although a solution may exist.
In particular, if a given mode \mode{} belongs to the schedule domains of two different applications, which have been independently scheduled in higher-priority modes, there is a risk of conflict as the inherited schedules may be incompatible.
We now formalize the notions of (virtual) legacy applications and conflicting modes. $\overline{X} = \appset \setminus X$ denotes the complement of $X$.
For each mode \modei, we define four application sets.
\begin{itemize}
 \item \textbf{Known applications} are the applications previously scheduled in higher-priority modes. The set of known applications of mode \modei is denoted by $\knownAppi = \cup_{j = 1}^{i-1} \; \modej$.
 \item \textbf{Free applications} are the newly scheduled applications in mode \modei, that is, no higher-priority mode belongs to the schedule domain of these applications. The set of free applications of mode \modei is denoted by $\freeAppi = \modei \; \cap \; (\appset \setminus \knownAppi) = \modei \; \cap \; \overline{\knownAppi}$.
 \item \textbf{Legacy applications} are the applications previously scheduled in higher-priority modes that must be scheduled in mode \modei. Since applications have a single schedule domain, \modei necessarily belongs to the same schedule domain as these higher-priority modes and the legacy application schedules must be inherited. The set of legacy applications of mode \modei is denoted by $\legAppi = \modei \; \cap \; {\knownAppi}$.
 \item \textbf{Virtual legacy applications} are the applications previously scheduled in higher-priority modes that are not scheduled in mode \modei. The set of virtual legacy applications of mode \modei is denoted by $\virtlegAppi = (\appset \setminus \modei)\; \cap \; {\knownAppi} = \overline{\modei} \; \cap \; {\knownAppi}$. The virtual legacy applications of \modei are not executed in \modei; they simply have been scheduled in higher-priority modes. As illustrated in \cref{exp:sched_conflict_basic}, it may be necessary to reserve the space for some of these virtual legacy applications to avoid future inheritance conflicts.
\end{itemize}

\squarepar{%
The schedules of two applications \appA and \appB are said to be in conflict if two tasks, one from \appA and one from \appB, are mapped to the same node and are scheduled in overlapping time intervals.
We denote by $s(\appA) \cap s(\appB) \not= \emptyset$ the property that \appA and \appB are in conflict.%
}

\begin{definition}[Conflict-free]
A set of applications $S$ is said to be conflict-free when there is no conflict between the schedules of the applications in $S$. We denote by \cf{S} the property that $S$ is conflict-free, and $\overline{\cf{S}}$ denotes that the set $S$ is in conflict.
Formally,
\begin{equation*}
\cf{S} \quad \Leftrightarrow  \quad \bigcap_{A \in S} \; s(A) = \emptyset
\end{equation*}
A mode is conflict-free if its legacy applications are conflict-free. In other words, $\forall \modei \in \modeset$,
\begin{equation*}
\cf{\modei} \quad \Leftrightarrow  \quad \cf{\legAppi}
\end{equation*}
The schedule \sched{\modei} of mode \modei is valid only if $\cf{\modei}$.
\end{definition}

\begin{corollary}
\label{cor:nec_precond}
	A valid schedule for mode $\modei \in \modeset$ can only exist if the legacy applications of \modei are conflict-free, that is,
	\begin{equation*}
	\cf{\legApp{i}} \; \Leftarrow \; \text{``Sched(\modei) is feasible''}
	\end{equation*}
\end{corollary}

\proof Using~\cref{exp:sched_conflict_basic} as a counter-example, $\overline{\cf{\legApp{4}}}$ makes it impossible to derive a valid schedule for \mode{4}.
\qed

\subsubsection{Minimal Inheritance Constraints}
\label{subsec:min_constraints}

\squarepar{%
The single-mode schedule synthesis is complete: If the problem is feasible for mode \modei , then \cref{alg:outerlayer} finds a valid schedule.
In particular, the scheduled mode is conflict-free, \cf{\modei}.
In the multi-mode case, certain applications are subject to continuity constraints, which are satisfied by fixing the schedules of legacy applications \legAppi in the MILP formulation for mode~\modei.
However, by \cref{cor:nec_precond}, this leads to a feasible schedule only if \cf{\legAppi}.%
}

\cref{exp:sched_conflict_basic} shows that inheriting legacy applications is not sufficient to prevent conflicts.
Thus, we now derive the subset of virtual legacy applications \virtlegAppi of mode \modei that is necessary and sufficient to reserve in order to guarantee the absence of a conflict due to continuity constraints.
More formally, our goal is
\begin{align}\label{eq:obj_conflict-free}
&\forall \, k \in [1..i-1], \; \text{``\sched{\mode{k}} is feasible''} \quad \Rightarrow \quad \cf{\legApp{i}}
\end{align}

To this end, we first formalize the constraints on the \sched{} function such that continuity constraints are enforced and conflicts are prevented as follows

\begin{align}
\label{eq:sched_redef}
Sched: \; &\modeset \; \longmapsto \; \sched{M}\\
\nonumber
s.t. \quad
		& \cf{\modei}\\
\nonumber
		& \forall\, \app \in \legApp{i} \cap \modej, \, j < i, \;	s_\modei(\app) = s_\modej(\app)\\
\nonumber
		& \forall\, \app \in \freeAppi, \; s(\app) \; \cap \; s(\minvirtlegApp{i}{\app}) = \emptyset
\end{align}

The first constraint in \cref{eq:sched_redef} ensures that the schedule is valid. The second one enforces the continuity constraints of applications. The third constraint enforces \cref{eq:obj_conflict-free} based on the idea that the free applications \freeAppi in mode \modei must be compatible with the schedules of some virtual legacy applications.
\cref{thm:minVirtLegacy} derives for any free application the minimal set of such virtual legacy applications. 

\begin{theorem}[Minimal virtual legacy sets]
\label{thm:minVirtLegacy}
For mode \modei and free application $\app \in \freeAppi$, the minimal set of virtual legacy applications \minvirtlegApp{i}{\app} necessary and sufficient to satisfy \eqref{eq:obj_conflict-free} is
\begin{equation}\label{eq:minVirtLegSets}
\minvirtlegApp{i}{\app} = \{ \appX \in \virtlegAppi \; |\;  \exists\, j>i, \quad \app \in  \legAppj \;\; \wedge \;\; \appX \in \legAppj \}
\end{equation}
\end{theorem}
\proof Please refer to \cref{app:proof} for the proof of this theorem.
\qed

\squarepar{%
Finally, we set the sum of all message deadlines, $\sum_{m_i\in \messageset} \, m_i.d$, as objective function of the MILP solver. Maximizing this allows to limit the constraints inherited between the different modes and improves the schedulability of the multi-mode schedule synthesis problem.%
}


\section{Experimental Evaluation}
\label{sec:ttw_evaluation_sched}
\vspace{-1mm}

Our experimental evaluation of \TTW answers the following questions:
\begin{itemize}
 \item \final{Are our timing and energy models of the \TTnet stack accurate (\cref{subsec:ttnet_eval})?}
 \item How big are the energy savings due to \TTnet's round-based communication (\cref{subsec:ttnet_eval})?
 \item Can the minimal inheritance constraints of \TTW's real-time scheduler effectively reduce the number of rounds while respecting the persistency of applications (\cref{subsec:benefits})?
 \item How long does it take to solve the multi-mode schedule synthesis problem (\cref{subsec:solving})?
\end{itemize}
\final{All the evaluation artifacts (implementation, raw data, and scripts) are openly available~\cite{repoTTW}.}

\vspace{-1mm}
\subsection{\TTnet Model Validation and Efficiency of Communication Rounds}
\label{subsec:ttnet_eval}

\squarepar{%
We begin by validating our \TTnet model and investigating the efficiency of rounds.
Before discussing our results, we detail the evaluation scenario and the design of our experiments.

\fakepar{Evaluation scenario}
Using 27 \DPP nodes on the FlockLab testbed~\cite{FlockLab}, we program \TTnet to execute one round of \nslots regular slots, followed by \nslots rounds with one regular slot each.
For each of these rounds, we collect the round length and the radio-on time.
Both values are measured in software using a 32\kHz timer, which gives a measurement resolution of about 30\us.
For $H=4$ and $N=2$, we test different number of slots per round \nslots and payload sizes $L$.
For each setting, we measure the round length and radio-on time experienced by each individual node in the network, and we compare the results with our \TTnet model.%
}

\afterpage{
\begin{table}
  \centering
  \caption{
  KPIs from the \TTnet model validation for different series and the corresponding model estimates of the round length \Tround and the energy savings $E$.
  \capt{
    {\ssymbol{2}}marks series in which, due to construction work taking place in the building where FlockLab is deployed, the number of collected metric values is insufficient to compute the KPIs; the reported values are then the maximum round length and the minimum energy saving across all runs in the series.
  }}
  \vspace{-0mm}
  \label{table:KPIs}
  {\smaller \input{\TablePath/KPIs.csv}}
\end{table}}

\squarepar{%
\fakepar{Experimental design}
We design our real-world experiments using \triscale~\cite{jacob2020TriScale}, a framework that facilitates the reproducibility of networking evaluations by allowing to make performance claims with quantifiable confidence.
\triscale distinguishes \emph{metrics}, which are computed over one run of the evaluation scenario, and \emph{key performance indicators} (KPIs), which capture the expected performance across a series of runs.
Concretely, KPIs are percentiles of the underlying distribution of the metric estimated with a certain level of confidence~\cite{jacob2020TriScale}.

We consider two performance dimensions: the empirical worst-case length of a round and the average per-node energy savings due to the use of rounds.
As we are interested in the worst-case, we take the maximum measurement across all nodes as metric for one run.
The true worst-case across any run is the 100th percentile of the metric distribution, which cannot be estimated with a finite number of runs.
We thus take as KPI the 95th percentile of our round time metric with a desired confidence level of 95\%.
For the average energy savings, we use the median values across nodes as metric, and estimate the 5th percentile of that metric with 95\% confidence.
\TriScale indicates that a minimum of 59 runs are needed for these estimations.
It has been shown that the experimental conditions on FlockLab exhibit seasonal components~\cite{jacobFlockLabLinkQuality}.
Therefore, to avoid any bias in our results, we perform the series of runs over a span of one week during which we randomly schedule 60 runs for each setting.
To investigate the reproducibility of the results, we perform 3 series of tests over the course of six months.
We test our \TTnet implementation with 5, 10, and 30 slots per round, and a payload of 8, 16, and 64 bytes.
This results in a total of 1620 individual runs.%
}

\begin{figure}[!tb]
  \centering
  \begin{subfigure}{0.48\linewidth}
    \centering
    \href{\ttwfig{Figure-8}}{%
    \includegraphics[scale=0.85]{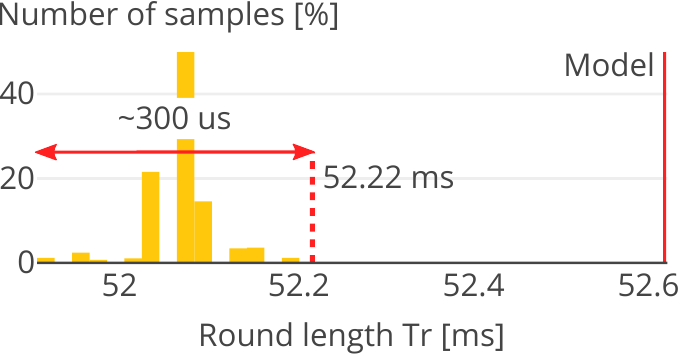}}
  \end{subfigure}
  \begin{subfigure}{0.48\linewidth}
      \centering
      \href{\ttwfig{Figure-8}}{%
      \includegraphics[scale=0.85]{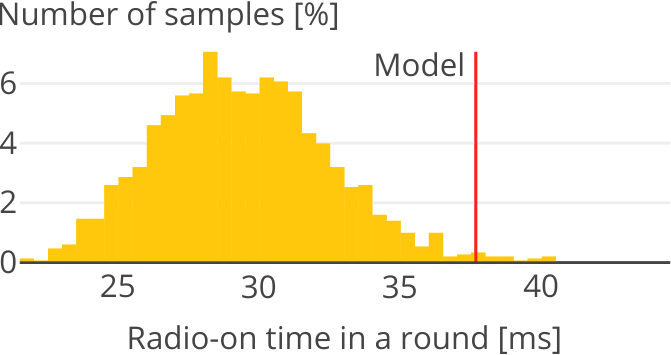}}
  \end{subfigure}
  \vspace{-2mm}
  \caption{Distributions of round length (left) and radio-on time (right) measurements from all 27 nodes on the FlockLab testbed, collected in one series of 60 runs with 5 slots per round and a payload size of 16\bytes.
  \capt{Our \TTnet models are accurate; the timing model is empirically safe.
  }}
  \vspace{-3mm}
  \label{fig:exampleSeries}
\end{figure}

\squarepar{%
\fakepar{Results---round length}
The experimental results in \cref{table:KPIs} confirm that our \TTnet model provides \final{a highly accurate estimate of} the length of a round.
The results are extremely stable across series: the largest difference in the KPI values corresponds to our measurement resolution of about 30\us.
Concretely, this means that the largest round length measured by any node is essentially the same.
Furthermore, the measured KPI values are very close to and consistently lower than the model estimates.
By definition of the KPI, we can claim that, with 95\% probability, at least 95\% of the runs of the evaluation scenario would yield a maximum round length that is less than or equal to the KPI value~\cite{jacob2020TriScale}, and thus smaller than the model estimate.
The left plot in \cref{fig:exampleSeries} shows the distribution of the round length measurements from all the nodes collected during one series of 60 runs. We observe that the distribution is very narrow (less than 300\us of spread); this is because all operations in a \TTnet round are time-triggered, so the measurement differences between nodes only come from the small differences in execution time of \baloo's end-of-round processing.%
}

\begin{remark}
  \squarepar{%
  The first series of test revealed a bug in the time synchronization software which led to an erratic behavior of certain nodes in some corner cases. This bug was fixed before the second series and the erratic data filtered out.
  In a previous version of this work~\cite{jacob2020Leveraging}, we presented a single case where one node measured a round time \emph{larger} than the model value. After closer investigation, it appears that this too was a consequence of the time synchronization bug, which we failed to detect and filter. The analysis script has been adapted accordingly; please refer to the TTW artifacts for more details~\cite{repoTTW}.%
  }
\end{remark}

\fakepar{Results---energy savings}
The results in \cref{table:KPIs} also show that the energy savings from the use of rounds in \TTnet are significant: the savings are around 30\% for a payload size of 16\bytes, which is representative of real-world \cps scenarios~\cite{hayat2016Survey,mager2019Feedback,preiss2017Crazyswarm}.
In general, the more slots are packed into a round (increasing \nslots) and the smaller the payload size (decreasing $L$), the higher the energy savings.
Since our energy KPI estimates the 5th percentile of the energy savings, we can claim that, with 95\% probability, at least 95\% of the runs in our evaluation scenario yield an average energy saving larger than or equal to the KPI value.
The right plot in \cref{fig:exampleSeries} shows the distribution of radio-on time measurements from all the nodes, collected in one series of 60 runs.
We see that the nodes experience significant differences in radio-on time during a round. This is expected since the nodes turn off their radio as soon as they have transmitted $N$ times during a Glossy flood, which happens earlier for nodes that are closer to the initiator of the flood.

\vspace{-3mm}
\subsection{Effectiveness of \TTW's Minimal Inheritance Approach}
\label{subsec:benefits}
\vspace{-1mm}

\squarepar{%
We now focus on \TTW's real-time scheduler and investigate whether the minimal inheritance constraints from \cref{subsec:min_constraints} help keep the total number of scheduled rounds low, while guaranteeing that persistent applications keep the same schedule across mode changes.%
}

\begin{figure}[!tb]
  \centering
  \includegraphics[width=0.5\linewidth]{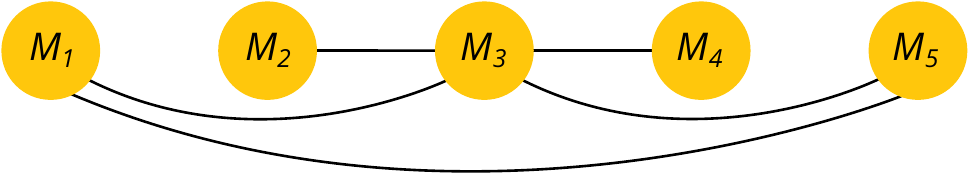}
  \vspace{-2mm}
  \caption{Mode graph used in the evaluation scenario of \cref{subsec:benefits}.}
  \vspace{-5mm}
  \label{fig:modeGraph_eval}
\end{figure}

\squarepar{%
\fakepar{Scenario}
We consider a scenario with 13 nodes running 15 different persistent applications.
There are in total 45 tasks and 30 messages.
The periods and deadlines vary between 10\s and 80\s.
The applications execute in 5 different modes; \cref{fig:modeGraph_eval} shows the mode graph with all possible transitions.
We synthesize schedules for the 5 modes on a standard laptop PC.
Besides our \emph{\final{minimal} inheritance} approach, we consider two baselines for comparison:
(\emph{i}) \emph{no inheritance}, which yields the minimum number of rounds under the (false) assumption that all applications are non-persistent; and (\emph{ii}) \emph{full inheritance}, which makes the (pessimistic) assumption that all applications executing in mode \modei are also executing in \modej.
In contrast to \emph{no inheritance}, the \emph{full inheritance} baseline guarantees continuity across mode changes but may find problems to be non-schedulable although a feasible solution exists.%
}

\begin{table}[t]
  \centering
  \caption{Solving times for each mode for three inheritance approaches \final{(in seconds)}}
   \vspace{-1mm}
  \label{table:solvingTimes}
  \final{\smaller
  \begin{tabular}{@{}l@{\qquad}c@{\qquad}c@{\qquad}c@{\qquad}c@{\qquad}c@{}}
    \toprule
    &\mode{1}&\mode{2}&\mode{3}&\mode{4}&\mode{5} \\
    \midrule
    \textbf{No inheritance}&5&$\approx$ 0&7&294&$\approx$ 0 \\
    \textbf{Minimal inheritance}&8&$\approx$ 0&$\approx$ 0&216&$\approx$ 0 \\
    \textbf{Full inheritance}&3&33&2&43&578 \\
    \bottomrule
  \end{tabular}%
  }
\end{table}

\begin{figure}[!tb]
  \href{\ttwfig{Figure-10}}{%
  \includegraphics[scale=0.8]{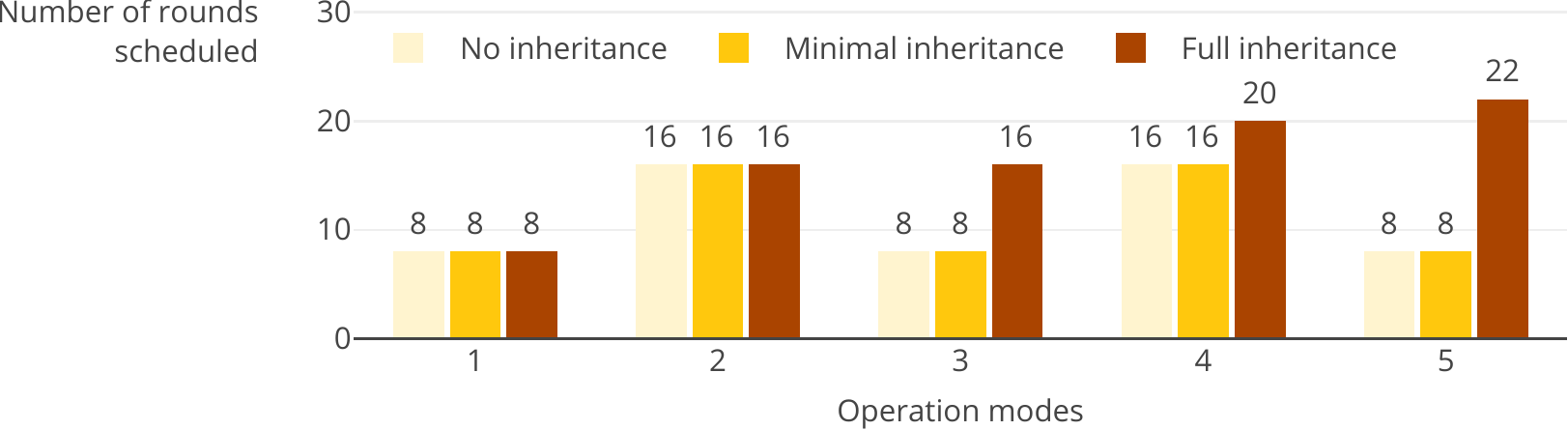}}
  \vspace{-2mm}
  \caption{Number of rounds scheduled in each mode for three inheritance approaches.
  \capt{We consider the number of rounds scheduled over 80\s, the least common multiple of the modes' hyperperiod.}}
  \vspace{-4mm}
  \label{fig:inheritance_eval}
\end{figure}

\fakepar{Results}
\cref{fig:inheritance_eval} shows for all approaches the number of rounds scheduled in each mode.
We can see that with \emph{full inheritance} the number of rounds steadily increases as it assumes that all previously scheduled application are still executing.
This not only wastes energy, it also limits scalability as the number of modes grows.
Our \emph{minimal inheritance} approach performs and scales significantly better as it considers only the relevant \final{constraints}.
In fact, in this particular scenario, \emph{minimal inheritance} performs optimally: it does not schedule more rounds than the absolute minimum, which is captured by the \emph{no inheritance} approach.

\vspace{-2mm}
\subsection{Offline Solving Time of \TTW's Real-Time Scheduler}
\label{subsec:solving}

Although \TTW targets \CPS scenarios in which the scheduling tables are synthesized before the system operation starts (\eg to a priori check that closed-loop stability can be guaranteed under given schedules~\cite{mager2019Feedback,baumann2019TCPS}), the solving time of the real-time scheduler is a relevant factor.

\fakepar{Scenario}
Using the scenario from the previous experiment, we measure the solving time for the three different inheritance approaches on a standard laptop PC.

\fakepar{Results}
We observe from \cref{table:solvingTimes} that the per-mode solving time ranges between less than a second and ten minutes, depending on the complexity of the mode.
For example, modes \mode{3} and \mode{4} contain the most applications, leading to more constraints in the formulation.
We also see that \emph{minimal inheritance} does not increase the overall solving time compared to \emph{no inheritance}.
This is because, by reserving some applications' schedule, certain problem variables are fixed, which reduces the computational load.
However, if too many variables are fixed, as shown by \emph{full inheritance}, the resulting problem may become harder to solve: more rounds are required, which may increases the number of variables and thus the solving time.


\vspace{-2mm}
\section{Conclusions}
\label{sec:ttw_conclusion}
\vspace{-1mm}

This paper presented \TTW, a time-triggered architecture for wireless \CPS.
\TTW provides guarantees on end-to-end deadlines by statically co-scheduling all tasks and messages in the system, while supporting runtime adaptability via mode changes that respect the schedules of persistent applications.
Our design of \TTW's real-time scheduler addressed key challenges concerning the formulation and tractability of the scheduling problem.
We leveraged synchronous transmissions to design a highly reliable, timing-predictable, and efficient low-power wireless communication stack that is robust to network dynamics.
We believe that \TTW takes wireless systems a major step closer the wired systems, which opens up several exciting opportunities for future \CPS applications that seemed so far out of reach.

\appendix

\section{Proof of \cref{thm:minVirtLegacy}}
\label{app:proof}

We first prove by recurrence that virtual legacy sets defined in \eqref{eq:minVirtLegSets} are \mbox{sufficient to satisfy \eqref{eq:obj_conflict-free}.}

For the highest priority mode \mode{1}, by definition, $\legApp{1} = \emptyset$, thus \cf{\legApp{1}}.
Let us assume that for any $k \in [1 .. i]$,
\sched{\mode{k}} is feasible in the sense of \eqref{eq:sched_redef}. This induces that \cf{S_k}, hence \cf{\legApp{k}} for any $k \in [1 .. i]$.
Let us finally assume that \legApp{i+1} is \emph{not} conflict-free; that is,
\begin{align}
\label{eq:square}
\overline{\cf{\legApp{i+1}}}  \quad
	\Leftrightarrow  \quad \bigcap_{A \in \legApp{i+1}} \; s(A) \not= \emptyset \quad
&
	\Rightarrow \quad \exists \, (A,B) \in \legApp{i+1}^2, \; s(A) \cap s(B) \not= \emptyset \\
\label{eq:square2}
	&
	\Rightarrow \quad
		\left\lbrace
		\begin{tabular}{@{\,}l@{\;\;}l@{\;\;}l}
		$\exists ! \, \mode{a}, \;A \in \freeApp{a} $&$\wedge $&$ a < i+1$ \\
		$\exists ! \, \mode{b}, \;B \in \freeApp{b} $&$\wedge $&$ b < i+1$ \\
		\end{tabular}
		\right.
\end{align}
where $\exists !$ means ``there exists a unique.''
Without loss of generality, we consider $a\leq b$. If $a=b$, then $S_a = S_b$ and $\cf{S_a} \equiv \cf{S_{b}}$. Therefore,
\begin{align}
	\bigcap_{A \, \in\,  S_a = S_b} \; s(A) = \emptyset
	\quad
	\Rightarrow
	\quad
	s(A) \cap s(B) = \emptyset
\end{align}
This contradicts \eqref{eq:square}, thus necessarily $a < b$; in other words, mode \mode{a} has higher priority than mode \mode{b}. Therefore, \app belongs either to \legApp{b} or \virtlegApp{b} by definition of those sets.
By hypothesis, \cf{S_{b}} and \eqref{eq:square2} : $B \in \freeApp{b}$, thus
\begin{equation*}
A \in \legApp{b} \quad \Rightarrow \quad s(A) \cap s(B) = \emptyset
\end{equation*}
which again contradicts \eqref{eq:square}. Hence necessarily,
$A \in \virtlegApp{b}$. Furthermore,
\begin{equation}
\left.
	\begin{tabular}{@{}l}
	$\eqref{eq:square2} : i+1 > b$\\
	$\eqref{eq:square} : A \in \legApp{i+1}$\\%
	$\eqref{eq:square} : B \in \legApp{i+1}$
	\end{tabular}
\right\rbrace
\; \text{Taking }b=i\text{ and }j = i+1, \quad \eqref{eq:minVirtLegSets}\; : \;
A \in \minvirtlegApp{b}{B}
\end{equation}
By hypothesis, \sched{\mode{b}} is feasible, thus $s(\appB) \; \cap \; s(\minvirtlegApp{b}{B}) = \emptyset$, which yields $s(B) \cap s(A) = \emptyset$ and contradicts \eqref{eq:square} again.
Therefore, the recurrence hypothesis is necessarily false.
Hence, if for any $k \in [1 .. i]$, \sched{\mode{k}} is feasible in the sense of \eqref{eq:sched_redef}, then \cf{\legApp{i+1}}.
By recurrence, we can conclude that the virtual legacy sets as defined by \eqref{eq:minVirtLegSets} are sufficient to satisfy \eqref{eq:obj_conflict-free}.

We now prove that the virtual legacy sets are also necessary.
Let us consider smaller virtual legacy sets than defined by \eqref{eq:minVirtLegSets}, that is, $\exists \; i \in [1..M],\, \app \in \freeAppi, \;\;\minminvirtlegApp{i}{A} \nsubseteq \minvirtlegApp{i}{A}$. Let us further assume that \sched{} is redefined to replace \minvirtlegApp{}{} by \minminvirtlegApp{}{}. By hypothesis,
\begin{align}
\label{eq:nec_cond1}
\exists \; \appX \in \appset, \; \appX \in \minvirtlegApp{i}{A}\;\wedge \; \appX
	& \notin \minminvirtlegApp{i}{A}\\
\label{eq:nec_cond4}
\text{Furthermore,} \quad \appX \in \minvirtlegApp{i}{A}
	&\quad \Rightarrow \quad
	\appX \in \virtlegApp{i} \quad \Rightarrow \quad {\appX \notin S_{i}}\\
\label{eq:nec_cond2}
\appX \in \minvirtlegApp{i}{A}
	&\quad \Rightarrow \quad
	\exists \; j>i, \quad \app \in \legApp{j} \; \wedge \; \appX \in \legApp{j}
\end{align}
Assuming that \sched{\modei} is feasible, the resulting schedule guarantees that
$\cf{\modei}$ and  $\forall \, \app \in \freeApp{i},\; s(\app) \, \cap \, s(\minminvirtlegApp{i}{A}) = \emptyset$.
However, \eqref{eq:nec_cond1} : $\appX \notin \minminvirtlegApp{i}{A}$ and \eqref{eq:nec_cond4} : $\appX \notin S_{i}$. Hence, schedule $s(\app)$ may be synthesized such that $s(A) \cap s(X) \not= \emptyset$. According to \eqref{eq:nec_cond2} : $(A,X) \in \legApp{j}^2$, which induces a conflict in mode \modej.
Hence, we can conclude that no sets \minminvirtlegApp{}{} smaller than \minvirtlegApp{}{} are sufficient to satisfy \eqref{eq:obj_conflict-free}.

\squarepar{%
Overall, this shows that the virtual legacy sets \minvirtlegApp{}{} as defined in \eqref{eq:minVirtLegSets} are both necessary and sufficient for the schedule synthesis method to satisfy \eqref{eq:obj_conflict-free}, that is, to guarantee that inheritance of schedules from legacy applications does not lead to conflicts in lower-priority modes.
In other words, \minvirtlegApp{}{} from \eqref{eq:minVirtLegSets} defines the sets of minimally restrictive constraints such that \sched{} as defined in \eqref{eq:sched_redef} satisfies \eqref{eq:obj_conflict-free}.
This completes the proof of \cref{thm:minVirtLegacy}.%
}

\bibliography{2020_ECRTS}
%

\end{document}